\documentclass{JFM-FLM_Au}
\IfFileExists{texmf/fonts/map/dvips/tex-gyre/qtm-ts1.map}{%
  \pdfmapfile{+texmf/fonts/map/dvips/tex-gyre/qtm-ts1.map}}{}

\usepackage{amsmath,amssymb,bm}
\usepackage{graphicx}
\usepackage{booktabs,tabularx,array}
\usepackage{xcolor}
   \defcitealias{caoshen2021}{CS21}
   \defcitealias{zhang2025windwave}{ZLD25}
\hypersetup{hidelinks}
\renewcommand{\Real}{\operatorname{Re}}
\renewcommand{\Imag}{\operatorname{Im}}

\newcolumntype{Y}{>{\raggedright\arraybackslash}X}

\newif\ifshowcomments\showcommentstrue

\title{An analytical framework for wave-coherent airflow over fast surface waves}
\author{Aditya K. Aiyer\aff{1}}
\affiliation{\aff{1}Department of Mechanical Engineering and Mechanics,
Lehigh University, Bethlehem, PA, USA, 18015}
\corresau{Aditya K. Aiyer, \email{aiyer@lehigh.edu}}
\lefttitle{A. Aiyer}
\righttitle{An analytical framework for wave-coherent airflow over fast surface waves}

\begin{document}
\maketitle
\begin{abstract}
\noindent
Classical theories attribute wind-wave momentum transfer to a critical layer, where the mean wind speed equals the wave phase speed. For waves that travel faster than the near-surface wind, this layer may lie outside the wave boundary layer or be absent altogether. We develop an analytical description of the wave-coherent airflow in this regime for a prescribed mean wind, using a Wentzel--Kramers--Brillouin--Jeffreys (WKBJ) approximation to the viscous Orr--Sommerfeld equation in wave-following coordinates. The solution combines an outer Rayleigh mode, which carries the dominant wave-coherent motion, with a thin viscous Stokes layer that supplies the phase shift producing the pressure form stress. At leading order, the form stress scales with the inverse square root of the wave Reynolds number, with a coefficient set by an integral of the outer response, and transfers momentum from the waves to the air. For an exponential mean profile, the WKBJ approximation gives closed-form expressions for the wave-coherent response. To accommodate the rapid near-surface variation of realistic wall-bounded profiles, we develop a long-wave expansion near the surface and match it to the outer WKBJ mode (LW--WKBJ). For wall-modelled simulations, whose first resolved level lies above
the viscous sublayer, we replace the molecular viscosity in the Stokes layer with a constant-stress effective viscosity. Using mean profiles from direct numerical simulations and wave-phase-resolved large-eddy simulations, the LW--WKBJ framework closely
reproduces the dominant wave-coherent velocity and pressure obtained
from numerical solutions and available simulation data. The predicted
form-stress coefficients also agree well with the available LES estimates.

\end{abstract}

\section{Introduction}
Momentum and energy transfer across the air--sea interface influences forecasts of weather and the wave field, and its parameterization is an important component of climate models \citep{janssen1991}. A central part of this exchange is carried by airflow motions that remain phase-locked to the surface waves. 
Locally generated wind waves are initiated by resonant forcing from turbulent pressure fluctuations \citep{phillips1957,lishen2025} and grow as momentum and energy are transferred from the wind \citep{plant1982,lishen2022}. Mature swell, by contrast, can propagate faster than the mean wind in the wave boundary layer, so that the wave-coherent stress transfers momentum from the waves to the wind  \citep{hanleybelcher2010}. The wave age $c/u_\ast$, where $c$ is the wave phase speed and $u_\ast$ is the air-side friction velocity, provides a useful, although not universal, measure of the transition between these regimes. 

Classical theories of wind-wave growth attribute the pressure
phase shift to a critical level, where the mean wind satisfies $\overline U=c$ \citep{miles1957}. Later work added viscous effects, wave-induced Reynolds stresses, and the feedback of
the growing waves on the mean wind
\citep{miles1959,miles1962,miles1993,janssen1991}. Analytical studies have since interpreted this critical-layer mechanism as an interaction between the surface wave and vorticity at the critical level \citep{carpenter2017} and related it to the Kelvin--Helmholtz and Holmboe instabilities of stratified shear flow \citep{kumarravichandran2026}. Its predictions have also been extended to coupled air--water profiles, water layers of finite depth and broader classes of mean profiles \citep{youngwolfe2014,kadam2023,bonfils2022,bonfils2023}. Field and laboratory observations support the
critical-layer mechanism for young and intermediate-age waves
\citep{hristov2003,carpenter2022}. For fast waves, this 
critical level
can lie outside the region dynamically coupled to the surface,
or be absent altogether, so a different mechanism must supply
the phase shift responsible for momentum transfer.

Field observations and wave-resolved simulations show that the wave-induced momentum flux can be directed upward: swell accelerates the near-surface airflow and can reduce or even reverse the total momentum flux 
\citep{smedman1994,grachevfairall2001,sullivan2008,Semedo2009,wu2016,wu2024review,aiyer2024}.  Boundary-layer models with a prescribed wave-induced stress reproduce this behavior and show that it can drive a low-level wind jet \citep{kudryavtsevmakin2004,hanleybelcher2008,ning2024parameterization}.  \citet{cohenbelcher1999} developed an asymptotic theory for turbulent flow over fast waves in which a mixing-length closure represents the wave-induced turbulent stress. Satellite observations \citep{ardhuin2009}, however, show that the attenuation of gentle swell is consistent with dissipation in a laminar air-side Stokes layer \citep{dore1978,perignon2014}, suggesting that molecular rather than turbulent stresses control the near-surface response.
Testing which stresses control the near-surface response requires the phase-resolved velocity and pressure close to the moving surface, which are difficult to measure in the field but are provided directly by wave-resolved simulations. 

Direct numerical simulations (DNS) and large-eddy simulations (LES) over slow and fast waves, both prescribed and dynamically evolving, have characterized the near-surface turbulence, pressure, and wave growth across wave ages \citep{sullivan2000,yangshen2010,haoshen2019,zhangdongshen2024,zhang2025windwave}. Their cost, however, restricts them to a limited range of wave ages and Reynolds numbers, which motivates reduced models of the wave-coherent airflow.

The classical reduced model linearizes the airflow about the mean wind, so that the wave-induced perturbation satisfies the viscous Orr--Sommerfeld equation. This formulation has described shear flow over wavy boundaries \citep{benjamin1959} and the initiation and growth of short, slow wind waves \citep{valenzuela1976,kawai1979,vangastel1985}. \citet{kumar2026} revisited this framework  using experimentally informed air- and water-side mean profiles, quantified its sensitivity to the friction velocity, surface drift and effective roughness, and used a perturbation-energy budget to identify tangential interfacial stress as the dominant transfer mechanism in the young-wave cases examined. \citet{muppirala2026} combined analytical Rayleigh solutions, Orr--Sommerfeld computations and asymptotic analyses to classify the instabilities that generate free-surface waves in curved shear flows. These approaches determine the complex phase speed, and hence the growth rate, as an eigenvalue of a coupled air--water stability problem.

Alternatively, the wave can be prescribed and the air-side response obtained from a forced boundary-value problem in wave-following coordinates. Combining LES with such linear viscous models, \citet{caodengshen2020} and \citet{caoshen2021} showed that the wave-coherent airflow over opposing and fast-following waves is governed primarily by wave kinematics, with the velocity and pressure response driven by the vertical motion of the surface. Wave-induced turbulent stresses remain important for the weaker components that control interfacial momentum transfer over opposing waves, but become subdominant to advection, pressure, and molecular viscosity at sufficiently high wave age, where the curvilinear model agrees closely with the simulations. \citet{narasimhan2026} subsequently generalized the formulation by retaining wave-coherent Reynolds-stress forcing, showing that these terms improve predictions over intermediate wave ages and that conventional real-valued mixing-length closures do not reproduce the magnitude and phase of the wave-induced stresses. Together, these studies identify a high-wave-age regime in which the molecular-viscosity Orr--Sommerfeld equation captures the leading phase-resolved response. 

In this paper we develop an analytical framework for the wave-coherent airflow over a prescribed fast wave. Our contribution is threefold. First, we construct a two-mode WKBJ solution of the molecular-viscosity Orr--Sommerfeld equation in wave-following coordinates, obtain an explicit leading-order expression for the pressure form stress, and introduce diagnostics, evaluated from the mean profile and wave parameters alone, that indicate where the outer approximation is admissible. The outer WKBJ approximation requires the mean profile to vary slowly over the decay length of the outer response, which holds for smooth shear layers but fails near the wall in wall-bounded flows. Second, we therefore represent the near-wall region by a long-wave expansion in powers of $k^2$ that retains the full variation of the mean velocity, and match it to the outer WKBJ mode. Third, we assess the resulting predictions in two stages: against numerical solutions of the viscous Orr--Sommerfeld equations for prescribed and simulation-derived mean profiles, which isolates the asymptotic error, and against wave-coherent fields from wall-resolved and wall-modelled LES, which tests the reduced dynamics.  The framework applies to the high wave age regime where a critical layer is absent in the wave boundary layer.

The paper is organized as follows. Section~\ref{sec:model} develops the asymptotic framework and its validity criteria.  Section~\ref{sec:test} first tests it with prescribed mean profiles and introduces the near-wall long-wave construction for wall-bounded profiles. It then applies the construction to the mean profiles of the wall-resolved LES of \citet[][hereafter CS21]{caoshen2021} and the coupled air--sea DNS of \citet[][hereafter ZLD25]{zhang2025windwave}. Section~\ref{sec:wrf_wall_modelled} presents wall-modelled simulations conducted with the Weather Research and Forecasting (WRF) model \citep{skamarock2019,zhu2023moving}, and \S\ref{sec:pressure_traction_coefficients} examines pressure and form stress for all cases. Conclusions follow in \S\ref{sec:conclusions}.
\section{Fast-wave asymptotics of the wave-following Orr--Sommerfeld equation}
\label{sec:model}

\subsection{Wave-following boundary-value problem}
\label{sec:wave_following_bvp}

We consider a prescribed, sufficiently smooth mean shear flow $\overline{U}$ over a
monochromatic deep-water wave propagating in the positive streamwise
direction. To pose the problem in a domain that conforms to the wavy surface, we introduce the wave-frame coordinate $\xi=x-ct$ and adopt the non-orthogonal
curvilinear transformation $\zeta=Z+G(\zeta)\eta(\xi)$
\citep{caoshen2021}. Here $Z$ is the physical height and $\zeta$ the mapped
height, both dimensional. The dimensionless function $G(\zeta)$ satisfies
$G(0)=-1$, so that the instantaneous surface $Z=\eta(\xi)$ maps to
$\zeta=0$. For the half-space formulation, $G$ is smooth and
$G(\zeta)\to0$ as $\zeta\to\infty$, so that the mapped coordinate
approaches the physical height far from the surface. The transformation
is assumed to remain invertible, which requires $1-\eta G'(\zeta)>0$, where
$G'=\mathrm dG/\mathrm d\zeta$. The governing equations are retained to
first order in the wave steepness, which is assumed small, $ak\ll1$, where $a$ is the wave amplitude, and $k=2\pi/\lambda$ is the wavenumber.
The surface elevation and linear orbital velocities are
\begin{equation}
\eta(\xi)=a\sin(k\xi), \qquad
\widetilde u_s(\xi)=akc\sin(k\xi), \qquad
\widetilde w_s(\xi)=-akc\cos(k\xi),
\label{eq:surface_kinematics}
\end{equation}
where \(c=(g/k)^{1/2}\) is the deep-water phase speed. Here and below,
the subscript $s$ denotes surface quantities.
For a wave-coherent quantity \(\widetilde f\), we adopt the Fourier
convention
\begin{equation}
\widetilde f(\xi,\zeta)
=
2\,\operatorname{Re}
\left\{
\widehat f(\zeta)\exp(\mathrm{i}k\xi)
\right\}.
\label{eq:fourier_convention}
\end{equation}
Phase-averaging the linearized momentum and continuity equations in these
coordinates reduces the problem to a single equation for the vertical-velocity amplitude \(\widehat w(\zeta)\) \citep{caoshen2021}:
\begin{equation}
-\frac{\nu}{\mathrm{i}k}
\left(
\frac{\mathrm{d}^2}{\mathrm{d}\zeta^2}-k^2
\right)^2
\widehat w
+
\left[
(\overline{U}-c)
\left(
\frac{\mathrm{d}^2}{\mathrm{d}\zeta^2}-k^2
\right)
-\frac{\mathrm{d}^2\overline{U}}{\mathrm{d}\zeta^2}
\right]
\widehat w
=
\nu\widehat\eta
\frac{\mathrm{d}^2}{\mathrm{d}\zeta^2}
\left(
G\frac{\mathrm{d}^2\overline{U}}{\mathrm{d}\zeta^2}
\right).
\label{eq:OS_dimensional}
\end{equation}
Here $\nu$ is the molecular kinematic viscosity of air.
The left-hand side is the Orr--Sommerfeld operator, comprising a viscous
term and the inviscid Rayleigh operator. The right-hand side is a geometric
forcing produced by the wave-following transformation.
Equation~\eqref{eq:OS_dimensional} omits the divergence of the wave-coherent Reynolds stresses, which is subdominant at the high wave ages considered here \citep{akervikvartdal2019,caoshen2021}. Note that, for wall-modelled LES, an effective viscosity replaces the molecular viscosity in \eqref{eq:OS_dimensional} and is introduced in \S\ref{sec:wrf_wall_modelled}.

Scaling lengths by \(k^{-1}\) and velocities by \(c\), we introduce the
nondimensional variables
\begin{equation}
z=k\zeta,
\qquad
\overline{u}(z)=\frac{\overline{U}(\zeta)}{c},
\qquad
w(z)=\frac{\widehat w(\zeta)}{c},
\qquad
R=\frac{c}{k\nu},
\label{eq:nondimensional_variables}
\end{equation}
in terms of which \eqref{eq:OS_dimensional} becomes
\begin{equation}
\frac{\mathrm{i}}{R}
\left(
w''''-2w''+w
\right)
+
(\overline{u}-1)
\left(
w''-w
\right)
-\overline{u}''w
=
\frac{1}{R}F(z).
\label{eq:OS_nondimensional}
\end{equation}
Bars denote mean quantities, tildes wave-coherent fields and hats their
dimensional Fourier amplitudes; nondimensional amplitudes are
unaccented, as in $w=\widehat w/c$. The symbol $G$ is
used for the mapping expressed in either coordinate, with its argument
identifying the representation. The geometric forcing is
\begin{equation}
F(z)
=
(k\widehat\eta)
\left\{
G(z)\overline{u}''(z)
\right\}''
=
-\frac{\mathrm{i}ak}{2}
\left\{
G(z)\overline{u}''(z)
\right\}''.
\label{eq:geometric_forcing}
\end{equation}
The wave Reynolds number $R=t_\nu/t_a$ is the ratio of the
molecular-diffusion time scale $t_\nu=(\nu k^2)^{-1}$ to the wave-advection
time scale $t_a=(kc)^{-1}$. For $R\gg1$, the small factor \(R^{-1}\)
multiplying the highest derivative makes \eqref{eq:OS_nondimensional} a
singularly perturbed fourth-order equation.
Under \eqref{eq:fourier_convention}, the surface kinematics
\eqref{eq:surface_kinematics} have the amplitudes
\begin{equation}
\widehat\eta=-\frac{\mathrm{i}a}{2},
\qquad
\widehat u_s=-\frac{\mathrm{i}akc}{2},
\qquad
\widehat w_s=-\frac{akc}{2}.
\label{eq:surface_fourier_amplitudes}
\end{equation}
We take the mean velocity to vanish at the mean interface,
\(\overline{u}(0)=0\), and define the nondimensional velocity defect
$\Delta(z)=1-\overline u(z)=[c-\overline U(\zeta)]/c$ and the signed
dimensional mean velocity relative to the wave,
$V(\zeta)=\overline U(\zeta)-c=-c\Delta(k\zeta)$.
The fast-wave regime is defined by $\Delta>0$, or equivalently $V<0$,
throughout the wave-coupled region, so that the Rayleigh operator
contains no critical-level singularity.
Imposing the kinematic and no-slip conditions at the surface,
$\widehat w(0)=\widehat w_s$ and $\widehat u(0)=\widehat u_s$, with the
amplitudes \eqref{eq:surface_fourier_amplitudes} and the linearized
continuity equation \eqref{eq:WKBJ_reconstruct_u} at $G(0)=-1$, gives
\begin{equation}
w(0)=w_s=-\frac{ak}{2},
\qquad
w'(0)=w_s'
=
\frac{ak}{2}
\left[
\overline{u}'(0)-1
\right].
\label{eq:surface_conditions}
\end{equation}
These are supplemented by the decay conditions $w,w'\to0$ as
$z\to\infty$. Defining the linear operator
\begin{equation}
\mathcal L_R[w]
\equiv
\frac{\mathrm{i}}{R}(D^2-1)^2w
+
(\overline{u}-1)(D^2-1)w
-\overline{u}''w,
\qquad
D\equiv\frac{\mathrm{d}}{\mathrm{d}z},
\label{eq:OS_operator}
\end{equation}
we decompose the wave-induced velocity into homogeneous and forced
responses \citep{akervikvartdal2019},
\begin{equation}
w=w_h+w_F,
\label{eq:homogeneous_forced_decomposition}
\end{equation}
where the homogeneous response \(w_h\) carries the surface conditions,
\begin{equation}
\begin{aligned}
\mathcal L_R[w_h]&=0,\\
w_h(0)&=w_s,
\qquad
w_h'(0)=w_s',
\qquad
w_h,w_h'\rightarrow0
\quad\text{as}\quad z\rightarrow\infty,
\end{aligned}
\label{eq:homogeneous_problem}
\end{equation}
and the geometrically forced correction \(w_F\) satisfies homogeneous
boundary conditions,
\begin{equation}
\begin{aligned}
\mathcal L_R[w_F]&=\frac{1}{R}F(z),\\
w_F(0)&=w_F'(0)=0,
\qquad
w_F,w_F'\rightarrow0
\quad\text{as}\quad z\rightarrow\infty.
\end{aligned}
\label{eq:forced_problem}
\end{equation}

\subsection{WKBJ construction}
\label{sec:outer_stokes_approximations}

Following the treatment of the Orr--Sommerfeld equation in
\citet{drazinreid2004} and the standard WKBJ procedure
\citep[][chap.~10]{benderorszag1999}, we write the homogeneous solution as
\begin{equation}
w_h(z)=\exp\left[\int^{z}q(s)\,\mathrm ds\right],
\qquad
q(z)=\frac{w_h'(z)}{w_h(z)} .
\label{eq:log_derivative_substitution}
\end{equation}
Using the identities $w_h''/w_h=q'+q^2$ and
$w_h''''/w_h=q'''+4qq''+3(q')^2+6q^2q'+q^4$, substituting
\eqref{eq:log_derivative_substitution} into \(\mathcal L_R[w_h]=0\) and
dividing by \(w_h\) reduces the linear fourth-order equation for \(w_h\)
to a nonlinear third-order equation for its logarithmic derivative \(q\):
\begin{align}
&-\Lambda^{-2}
\left[
q'''+4qq''+3(q')^2+6q^2q'+q^4-2q'-2q^2+1
\right]
\nonumber\\
&\hspace{2cm}
+(\overline{u}-1)\left(q'+q^2-1\right)-\overline{u}''=0,
\label{eq:q_equation_exact}
\end{align}
where $\Lambda=\left(\mathrm{i}R\right)^{1/2}$, so that
$\mathrm{i}/R=-\Lambda^{-2}$. This transformation is useful
because the asymptotic balance now depends directly on the magnitude
of \(q\). In the limit \(R\gg1\), distinct scalings of \(q\) emerge,
corresponding to solutions that vary on the outer \(O(1)\) scale and
on the faster \(O(R^{-1/2})\) viscous scale.
We therefore expand the logarithmic derivative as
\begin{equation}
q(z)
\sim
\Lambda \chi_0(z)
+
\chi_1(z)
+
\Lambda^{-1}\chi_2(z)
+\cdots .
\label{eq:q_expansion}
\end{equation}
Substituting into \eqref{eq:q_equation_exact} and collecting the
leading \(O(R)\) terms gives
\begin{equation}
\chi_0^2
\left[
(\overline{u}-1)-\chi_0^2
\right]
=0.
\label{eq:q0_eikonal}
\end{equation}
Equation~\eqref{eq:q0_eikonal} admits two distinct possibilities:
\(\chi_0\neq0\), which produces rapidly varying viscous modes, and
\(\chi_0=0\), for which the leading balance reduces to the outer Rayleigh
problem. Each case yields two independent solutions, which together
account for the four solutions of the fourth-order Orr--Sommerfeld
equation.

\subsubsection{Viscous Stokes modes}
\label{sec:viscous_modes}

We use the subscript $o$ for the outer Rayleigh branch and $\nu$ for
the viscous Stokes branch.
For the viscous modes, \(\chi_0\neq0\), and
\eqref{eq:q0_eikonal} gives
$\chi_0^2=\overline{u}-1=-\Delta$, so that 
$\chi_0
=
\pm \mathrm{i}\Delta^{1/2}$.
Since \(\Delta>0\), the viscous roots never vanish, and the viscous modes therefore contain no turning point.
Collecting the $O(R^{1/2})$ terms in
\eqref{eq:q_equation_exact} and using $\chi_0^2=\overline{u}-1$ gives
\begin{equation}
-2\chi_0^3\chi_1
-
5\chi_0^2\chi_0'
=0.
\label{eq:viscous_transport}
\end{equation}
This transport equation determines the slowly varying WKBJ
prefactor through
\begin{equation}
\chi_1
=
-\frac{5}{2}\frac{\chi_0'}{\chi_0} =-\frac{5}{4}\frac{\Delta'}{\Delta} .
\label{eq:viscous_q1_general}
\end{equation}
Truncating the expansion after \(\chi_1\) gives the WKBJ approximation
$q_\nu^{\mathrm{WKBJ}}=\Lambda\chi_0+\chi_1$ to the viscous logarithmic
derivative. The corresponding WKBJ mode \(W_\nu\), defined by
$W_\nu'/W_\nu=q_\nu^{\mathrm{WKBJ}}$, is
\begin{equation}
W_\nu(z)
=
W_\nu(0)
\exp\left[
\Lambda\int_0^z \chi_0(s)\,\mathrm{d}s
\right]
\exp\left[
\int_0^z \chi_1(s)\,\mathrm{d}s
\right].
\label{eq:viscous_solution_intermediate}
\end{equation}
The first exponential determines the rapid variation of the viscous
mode, with
\begin{equation}
    \Lambda \chi_0
=
(\mathrm{i}R)^{1/2}
\left(\pm \mathrm{i}\Delta^{1/2}\right)
=
\pm
\mathrm{e}^{3\mathrm{i}\pi/4}
R^{1/2}\Delta^{1/2}.
\end{equation}
The second exponential determines the slowly varying amplitude:
\begin{equation}
\exp\left[
\int_0^z \chi_1(s)\,\mathrm{d}s
\right]
=
\exp\left[
-\frac{5}{4}
\int_0^z
\frac{\Delta'(s)}{\Delta(s)}
\,\mathrm{d}s
\right]
=
\left[
\frac{\Delta(z)}{\Delta(0)}
\right]^{-5/4}.
\end{equation}
Normalizing each mode to unit surface amplitude, $W_{\nu,\pm}(0)=1$,
gives the two viscous WKBJ modes
\begin{equation}
W_{\nu,\pm}(z)
=
\left[
\frac{\Delta(z)}{\Delta(0)}
\right]^{-5/4}
\exp
\left[
\pm
\mathrm{e}^{3\mathrm{i}\pi/4}
R^{1/2}
\int_0^z
\Delta^{1/2}(s)\,\mathrm{d}s
\right].
\label{eq:viscous_modes}
\end{equation}

Since $\mathrm{e}^{3\mathrm{i}\pi/4}$
has a negative real part, the \(+\) branch decays away from the surface,
whereas the \(-\) branch grows exponentially. We therefore retain
$W_\nu(z)\equiv W_{\nu,+}(z)$
as the physically admissible viscous mode.

\subsubsection{Outer Rayleigh modes}
\label{sec:outer_modes}

For the second pair of solutions, $\chi_0=0$, so the $O(R^{1/2})$
contribution to the logarithmic derivative vanishes and
$q=\chi_1+O(R^{-1/2})=O(1)$. In this case, the
viscous terms in \eqref{eq:q_equation_exact} are smaller by
\(O(R^{-1})\), and the leading balance reduces to
\begin{equation}
q_o'+q_o^2=Q(z),
\qquad
Q(z)\equiv
1-\frac{\overline{u}''(z)}{\Delta(z)}.
\label{eq:outer_riccati}
\end{equation}
Writing $q_o=w_o'/w_o$ gives $q_o'+q_o^2=w_o''/w_o$, so
\eqref{eq:outer_riccati} is equivalent to the leading-order homogeneous
Rayleigh equation
\begin{equation}
w_o''-Q(z)w_o=0,
\label{eq:outer_rayleigh_equation}
\end{equation}
where \(w_o\) denotes an exact solution.
Thus, unlike the viscous mode, the leading outer solution is
independent of \(R\) and is determined entirely by the effective
Rayleigh potential \(Q(z)\). For a slowly varying \(Q\), we seek WKBJ approximations
\(W_{o,\pm}\) to the exact Rayleigh solutions \(w_o\). Neglecting
\(q_o'\) at leading order gives
$q_{o,\pm}^{(0)}=\pm Q^{1/2}$.
Retaining the leading correction due to the spatial variation of \(Q\) gives
\begin{equation}
q_{o,\pm}^{\mathrm{WKBJ}}
\equiv
\frac{W_{o,\pm}'}{W_{o,\pm}}
=
\pm Q^{1/2}
-\frac{1}{4}\frac{Q'}{Q}
+\cdots .
\label{eq:outer_q_WKBJ}
\end{equation}
For fast waves $Q(z)>0$
throughout the wave-coupled region and 
$Q^{1/2}$ is real and nonzero, so the two leading solutions vary
exponentially.
Integrating \eqref{eq:outer_q_WKBJ} gives the two outer WKBJ modes
\begin{equation}
W_{o,\pm}(z)
=
\left[
\frac{Q(z)}{Q(0)}
\right]^{-1/4}
\exp\left[
\pm
\int_0^z
Q(s)^{1/2}\,\mathrm{d}s
\right],
\label{eq:outer_WKBJ_modes}
\end{equation}
where both modes are normalized to unity at the surface. For \(Q>0\),
the positive-sign solution grows with height, whereas the negative-sign
solution decays. The physically admissible outer mode is therefore
$W_o\equiv W_{o,-}$.

\subsubsection{Validity of the outer WKBJ approximation}
\label{sec:outer_WKBJ_validity}
 Hereafter, we omit the branch subscripts and write
$W_o=W_{o,-}$ and $W_\nu=W_{\nu,+}$ for the decaying outer
and viscous modes, respectively. 
We now assess how closely the decaying WKBJ mode \(W_{o}\) in
\eqref{eq:outer_WKBJ_modes} approximates an exact solution \(w_o\) of
\eqref{eq:outer_rayleigh_equation}.
For locally constant \(Q>0\), the exact decaying Rayleigh solution is
$w_o(z)\propto\exp(-Q^{1/2}z)$, with local e-folding length
$\ell_Q=Q^{-1/2}$, or $k^{-1}Q^{-1/2}$ in dimensional terms.
The outer WKBJ approximation assumes that \(Q\) varies slowly on this
scale. The fractional change in \(Q\) over one local decay length is estimated by
\begin{equation}
\frac{\lvert Q'\rvert\ell_Q}{Q}
=
\frac{\lvert Q'\rvert}{Q^{3/2}}
\equiv
\epsilon_Q.
\label{eq:outer_slow_variation}
\end{equation}
Thus, \(\epsilon_Q\ll1\) means that the coefficient governing the
local decay of the outer mode remains nearly constant over the
distance on which the mode decreases by a factor of \(\mathrm e\).
Unlike the viscous-mode expansion, the accuracy of the outer WKBJ
approximation is controlled by the spatial variation of \(Q\) rather
than by \(R\). 
A direct measure of the approximation error is obtained by
substituting the decaying WKBJ mode \(W_{o}\) from
\eqref{eq:outer_WKBJ_modes} into the exact outer Rayleigh operator.
The resulting normalized residual is
\begin{equation}
\mathcal{R}_Q
\equiv
\frac{
W_{o}''-QW_{o}
}{
QW_{o}
}
=
\frac{5(Q')^2}{16Q^3}
-
\frac{Q''}{4Q^2}.
\label{eq:outer_WKBJ_residual}
\end{equation}
 The first contribution to
\(\mathcal{R}_Q\) may be written as
\begin{equation}
\frac{5(Q')^2}{16Q^3}
=
\frac{5}{16}\epsilon_Q^2,
\end{equation}
while the second contribution measures the local curvature of \(Q\).
For a smoothly varying profile characterized by a single vertical
length scale, the two contributions have the same asymptotic order.
Accordingly, \(\epsilon_Q\ll1\) provides a convenient leading
slow-variation criterion, while
\(\lvert\mathcal{R}_Q\rvert\ll1\) directly measures how closely the
WKBJ mode satisfies the outer Rayleigh equation.

\subsection{Surface matching and the composite solution}
\label{sec:surface_matching}

Imposing decay as $z\to\infty$ leaves the second-order outer
Rayleigh problem with a single free amplitude. 
The thin viscous Stokes mode provides the second degree of freedom, and the homogeneous response is written as
\begin{equation}
W_h(z)
=
A_oW_o(z)
+
A_\nu W_\nu(z),
\label{eq:homogeneous_response}
\end{equation}
where both decaying modes are normalized so that
$W_o(0)=W_\nu(0)=1$.
The surface conditions \eqref{eq:surface_conditions} then give
\begin{equation}
A_o+A_\nu=W_h(0) = w_s,
\qquad
A_oW_o'(0)+A_\nu W_\nu'(0)=W_h'(0) = w_s',
\label{eq:surface_matching_system}
\end{equation}
and hence
\begin{equation}
A_\nu
=
\frac{
w_s'-w_sW_o'(0)
}{
W_\nu'(0)-W_o'(0)
},
\qquad
A_o
=
\frac{
w_sW_\nu'(0)-w_s'
}{
W_\nu'(0)-W_o'(0)
}.
\label{eq:matched_amplitudes}
\end{equation}

To determine the leading structure of the composite solution, we next
estimate the relative magnitudes of \(A_o\) and \(A_\nu\) in
\eqref{eq:matched_amplitudes}. These estimates apply as $R\to\infty$
with the nondimensional mean profile and mapping held fixed. We assume that $\Delta$ is bounded away from zero and that the mean-profile and mapping derivatives are bounded independently of $R$. The normalized outer mode then has an $O(1)$ surface slope, $W_o'(0)=O(1)$.
The prescribed surface quantities \(w_s\) and \(w_s'\) are then both
\(O(ak)\), so the scaling of \(A_\nu\) is set by the difference between
the surface gradients of the two modes. From \eqref{eq:outer_WKBJ_modes}
and \eqref{eq:viscous_modes},
\begin{align}
W_o'(0)&=-Q(0)^{1/2}-\frac{Q'(0)}{4Q(0)},
\label{eq:outer_surface_derivative}\\
W_\nu'(0)&=\mathrm{e}^{3\mathrm{i}\pi/4}R^{1/2}\Delta(0)^{1/2}
-\frac{5}{4}\frac{\Delta'(0)}{\Delta(0)}.
\label{eq:stokes_surface_derivative}
\end{align}
The outer slope is independent of $R$ for a fixed profile, whereas the
Stokes mode changes by \(O(1)\) across a layer of thickness
\(O(R^{-1/2})\), so that $W_\nu'(0)=O(R^{1/2})$. The numerator of
$A_\nu$ is therefore \(O(ak)\), while
\(W_\nu'(0)-W_o'(0)=O(R^{1/2})\). Hence
\begin{equation}
A_\nu=O(akR^{-1/2}),
\qquad
A_o=w_s-A_\nu
=
-\frac{ak}{2}+O(akR^{-1/2}).
\label{eq:mode_amplitude_scaling}
\end{equation}
The outer mode sets the leading \(O(ak)\) velocity response,
while the Stokes mode provides a smaller \(O(R^{-1/2})\) correction.
Its complex phase, however, supplies the leading phase shift required
for form stress.

These estimates also require the outer and viscous scales to remain
distinct,
\begin{equation}
Q^{1/2}\ll(R\Delta)^{1/2},
\label{eq:mode_separation}
\end{equation}
so that the local Rayleigh decay length greatly exceeds the Stokes-layer
thickness. They are not uniform over families whose nondimensional
shear-layer thickness decreases with $R$.
The final construction requires $\Delta>0$ and $Q>0$ over the wave-coupled region,
small values of $\epsilon_Q$ and $|\mathcal R_Q|$, and the scale separation
\eqref{eq:mode_separation}. Each condition can be evaluated from the
mean profile and wave parameters before the solution is computed.

The geometric forcing \eqref{eq:geometric_forcing} enters \eqref{eq:OS_nondimensional} with an explicit factor $R^{-1}$. In the outer region, where the viscous term is subleading, the forced response satisfies $w_F''-Qw_F=-F/(R\Delta)$. For a smooth mean profile $F=O(ak)$, so $w_F=O(ak/R)$, which is smaller than the $O(akR^{-1/2})$ Stokes correction from surface matching. Under the present sine convention, $F$ is purely imaginary and the leading outer operator is real, so $w_F$ is also imaginary to leading order. Through the pressure reconstruction, it produces a real pressure correction of $O(R^{-1})$, one order beyond the leading $O(R^{-1/2})$ pressure form stress considered here. The forced response is small for all the mean profiles considered here \citep[see also][]{caoshen2021}: for each case with a numerical reference, the numerical solution retains $F$, and comparison with the corresponding homogeneous solution confirms that its effect is small (\S\ref{sec:test} and~\S\ref{sec:pressure_traction_coefficients}).

\subsection{Velocity and pressure reconstruction}
\label{sec:velocity_pressure_reconstruction}
Once the matched WKBJ solution is known, the wave-induced streamwise
velocity and pressure follow from the linearized continuity and vertical
momentum equations. Returning to dimensional variables, we use
$\widehat w(\zeta)\simeq c\,W_h(k\zeta)$ at the retained
order. Continuity in the wave-following coordinates gives
\begin{equation}
\widehat u(\zeta)
=-G(\zeta)\widehat\eta\,\frac{\mathrm d\overline{U}}{\mathrm d\zeta}
+\frac{\mathrm i}{k}\frac{\mathrm d\widehat w}{\mathrm d\zeta}.
\label{eq:WKBJ_reconstruct_u}
\end{equation}
Under the sine convention $\widehat\eta=-\mathrm i a/2$, the real and imaginary parts are
\begin{align}
\operatorname{Re}\{\widehat u\}
&=-\frac{1}{k}\frac{\mathrm d}{\mathrm d\zeta}
  \operatorname{Im}\{\widehat w\},
\label{eq:WKBJ_reconstruct_u_real}\\
\operatorname{Im}\{\widehat u\}
&=\frac{1}{k}\frac{\mathrm d}{\mathrm d\zeta}
  \operatorname{Re}\{\widehat w\}
  +\frac{a}{2}G(\zeta)\frac{\mathrm d\overline{U}}{\mathrm d\zeta}.
\label{eq:WKBJ_reconstruct_u_imag}
\end{align}
With constant air density $\rho_a$, the dimensional pressure amplitude
satisfies the vertical momentum balance
\begin{equation}
\frac{1}{\rho_a}\frac{\mathrm d\widehat p}{\mathrm d\zeta}
=-\mathrm i k(\overline{U}-c)\widehat w
+\nu\left(\frac{\mathrm d^2}{\mathrm d\zeta^2}-k^2\right)\widehat w.
\label{eq:WKBJ_pressure_gradient}
\end{equation}
Integrating from $\zeta$ to the upper boundary $H$, where
$\widehat p(H)=0$, yields
\begin{equation}
\frac{\widehat p(\zeta)}{\rho_a}
=\mathrm i k\int_\zeta^H[\overline{U}(s)-c]\widehat w(s)\,\mathrm ds
-\nu\int_\zeta^H
\left(\frac{\mathrm d^2}{\mathrm ds^2}-k^2\right)
\widehat w(s)\,\mathrm ds.
\label{eq:WKBJ_reconstruct_p}
\end{equation}
The value of $H$ is specified for each comparison; the half-space
expression follows by letting $H\to\infty$ and imposing pressure decay. Separating \eqref{eq:WKBJ_reconstruct_p} into
real and imaginary parts gives
\begin{align}
\frac{\operatorname{Re}\{\widehat p(\zeta)\}}{\rho_a}
&=-k\int_\zeta^H[\overline{U}(s)-c]\operatorname{Im}\{\widehat w(s)\}\,\mathrm ds
\nonumber\\
&\quad-\nu\int_\zeta^H
\left(\frac{\mathrm d^2}{\mathrm ds^2}-k^2\right)
\operatorname{Re}\{\widehat w(s)\}\,\mathrm ds,
\label{eq:WKBJ_reconstruct_p_real}\\
\frac{\operatorname{Im}\{\widehat p(\zeta)\}}{\rho_a}
&=k\int_\zeta^H[\overline{U}(s)-c]\operatorname{Re}\{\widehat w(s)\}\,\mathrm ds
\nonumber\\
&\quad-\nu\int_\zeta^H
\left(\frac{\mathrm d^2}{\mathrm ds^2}-k^2\right)
\operatorname{Im}\{\widehat w(s)\}\,\mathrm ds.
\label{eq:WKBJ_reconstruct_p_imag}
\end{align}
The velocity derivatives in these expressions are evaluated directly
from the WKBJ modes. The physical wave-coherent fields are then recovered as
\begin{equation}
\widetilde f(\xi,\zeta)
=2\operatorname{Re}\{\widehat f(\zeta)\}\cos(k\xi)
-2\operatorname{Im}\{\widehat f(\zeta)\}\sin(k\xi),
\qquad f\in\{u,p\}.
\label{eq:WKBJ_reconstruct_physical_fields}
\end{equation}

\section{Assessment with prescribed and simulation-derived mean profiles}
\label{sec:test}

The framework of \S\ref{sec:model} requires only the surface elevation and the mean wind profile. We first assess it with two prescribed profiles: an exponential profile, whose WKBJ phase integrals have closed forms (\S\ref{sec:exp}), and a shifted logarithmic profile, whose rapid near-surface variation motivates the near-wall long-wave construction (\S\ref{sec:log_longwave}). We then apply the construction to the mean profiles of the wall-resolved LES of \citetalias{caoshen2021} and the coupled air--sea DNS of \citetalias{zhang2025windwave} (\S\ref{sec:simulation_means}). For the prescribed profiles and the DNS mean, the reference is the numerical solution of the wave-following Orr--Sommerfeld equation; for \citetalias{caoshen2021} it is the wave-coherent LES field. Table~\ref{tab:test_cases} summarizes all cases, including the wall-modelled simulations of \S\ref{sec:wrf_wall_modelled}. Unless stated otherwise, discrepancies are relative $L_2$ norms over height, normalized by the reference; complex values use the complex field, and component-wise values normalize each real or imaginary part by its own norm.

\begin{table}
\centering
\small
\begin{tabular}{llll}
\hline
Case & Mean profile & Reference & Inner treatment  \\
\hline
Exponential (\S\ref{sec:exp}) & Analytic & Numerical solution & WKBJ from the surface \\
Shifted log (\S\ref{sec:log_longwave}) & Analytic & Numerical solution & LW--WKBJ \\
\citetalias{caoshen2021} (\S\ref{sec:cao_wrles}) & Wall-resolved LES & LES harmonics & LW--WKBJ \\
\citetalias{zhang2025windwave} (\S\ref{sec:zhang_swell}) & DNS & Numerical solution & LW--WKBJ \\
WRF (\S\ref{sec:wrf_wall_modelled}) & Wall-modelled LES & LES harmonics & LW--WKBJ \\
\hline
\end{tabular}
\caption{ All test cases use the molecular viscous mode except WRF,
which uses the effective-viscosity closure of \S\ref{sec:wrf_wall_modelled}.
Numerical solutions are finite-difference (FD) or Chebyshev solutions of
the wave-following Orr--Sommerfeld equation with the mean profile either prescribed analytically or obtained from simulations.}
\label{tab:test_cases}
\end{table}

\subsection{Exponential shear profiles}
\label{sec:exp}
The exponential mean profile provides a controlled benchmark for the
fast-wave theory because it allows the wind speed and shear-layer thickness
to be varied independently, while retaining a closed-form expression for
each WKBJ phase integral. We prescribe
\begin{equation}
\overline{U}(\zeta)=U_0\left(1-\mathrm{e}^{-\zeta/\delta}\right),
\qquad
\overline{u}(z)=B\left(1-\mathrm{e}^{-z/d}\right),
\label{eq:expprofile}
\end{equation}
where $U_0$ is the asymptotic wind speed, $\delta$ is the shear-layer
thickness, $B=U_0/c$, and $d=k\delta=2\pi\delta/\lambda$. With $A=1-B$,
the velocity defect is $\Delta(z)=A+B\mathrm{e}^{-z/d}$ and the Rayleigh
potential is $Q(z)=1+B\mathrm{e}^{-z/d}/[d^2\Delta(z)]$. When the
asymptotic wind speed is below the wave phase speed, the velocity defect
stays positive at all heights and the
profile doesn't contain a critical level. 
WKBJ accuracy additionally requires $\Delta$ and $Q$ to vary slowly. At fixed wind-speed ratio, this variation is controlled by $d$: decreasing $d$ confines the mean shear and curvature to a thinner near-surface region, providing a more stringent test of the approximation.

The substitutions $y=\Delta^{1/2}$ and $l=Q^{1/2}$ 
reduce both phase
integrals to elementary integrals, giving
\begin{align}
\Phi(z)
&\equiv\int_0^z\sqrt{\Delta(s)}\,\mathrm{d}s
\nonumber\\
&=2d\left[1-y(z)\right]+\sqrt{A}\,z
 +2d\sqrt{A}\ln\left[
 \frac{y(z)+\sqrt{A}}{1+\sqrt{A}}
 \right],
\label{eq:Phi}\\
\Psi(z)
&\equiv\int_0^z\sqrt{Q(s)}\,\mathrm{d}s
\nonumber\\
&=z+2d\ln\left[\frac{l(z)+1}{l(0)+1}\right]
 +2\sqrt{1+d^2}\ln\left[
 \frac{l(0)+\sqrt{1+d^{-2}}}{l(z)+\sqrt{1+d^{-2}}}
 \right]
\nonumber\\
&\quad+
 \left(d-\sqrt{1+d^2}\right)\ln\Delta(z).
\label{eq:Psi}
\end{align}
Both phases vanish at the surface, and differentiation confirms that
\(\Phi'=\sqrt{\Delta}\) and \(\Psi'=\sqrt{Q}\).
The normalized, decaying modes therefore become
\begin{equation}
\begin{aligned}
W_\nu(z)
&=\Delta(z)^{-5/4}
\exp\left[\mathrm{e}^{3\mathrm{i}\pi/4}\sqrt{R}\,\Phi(z)\right],
\\
W_o(z)
&=\left[\frac{Q(z)}{Q(0)}\right]^{-1/4}
\exp\left[-\Psi(z)\right].
\end{aligned}
\label{eq:exp_WKBJ_modes}
\end{equation}
Together with the matched amplitudes
\eqref{eq:matched_amplitudes}, these expressions give the homogeneous
WKBJ response without numerical evaluation of the phase integrals.
The phase integrals are exact, while the modes themselves remain
asymptotic approximations to the Orr--Sommerfeld solution.

\begin{figure}[htbp]
\centering
\includegraphics[width=\textwidth]{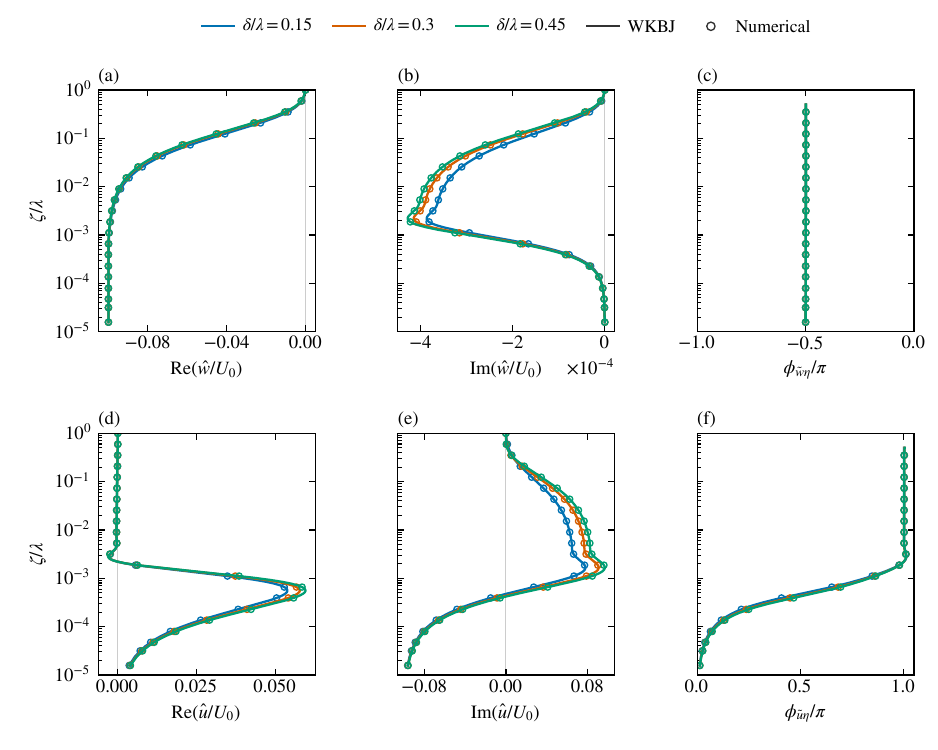}
\caption{Wave-coherent velocity for the exponential mean profile at
$U_0/c=0.5$, $R=10^5$ and $ak=0.1$, with $\delta/\lambda=0.15$, $0.30$
and $0.45$ (blue, orange and green). Rows show vertical and streamwise
velocity; columns show the real part, imaginary part and phase relative
to the surface elevation, $\phi_{\tilde f\eta}=\arg(\widehat f\widehat\eta^*)$.
Velocities are normalized by $U_0$. Solid lines show the homogeneous WKBJ
approximation and open circles the finite-difference solution including
mapped forcing. The numerical domain has height $H=\lambda$.}
\label{fig:exp_sweep_delta}
\end{figure}

We first consider profiles for which these conditions are well satisfied.
Figure~\ref{fig:exp_sweep_delta} shows the response at $R=10^5$ and
$ak=0.1$ for $\delta/\lambda=0.15$, $0.30$ and $0.45$ at fixed
$U_0/c=0.5$, with the wave speed and wavenumber unchanged. For these
profiles, and for a companion sweep $0.1\leq U_0/c\leq0.9$ at
$\delta/\lambda=0.25$ (not shown), $\epsilon_Q\leq0.17$,
$|\mathcal R_Q|\leq0.02$ and $[Q/(R\Delta)]^{1/2}\leq0.01$ at every
height, so the conditions summarized in \S\ref{sec:surface_matching} hold
throughout the domain. Under these conditions the geometric forcing is
also negligible: the full and homogeneous numerical solutions differ by
less than $0.001\,\%$ in vertical velocity, streamwise velocity and
pressure.

The numerical reference solves the wave-following Orr--Sommerfeld
equation on $0\leq\zeta\leq H$, with $H=\lambda$ and
$\widehat w=\widehat w'=0$ at the upper boundary. On this finite domain
we use $G(\zeta)=\zeta/H-1$, or $G(z)=z/(kH)-1$, which maps the wavy
lower boundary to $\zeta=0$ and leaves the upper boundary fixed at
$\zeta=H$. 
A nonuniform grid with 4001 points is refined at both
boundaries to resolve the viscous layers. The WKBJ calculation uses the
analytical mode derivatives. 

Increasing $\delta/\lambda$ from $0.15$ to $0.45$ leaves the dominant real
vertical-velocity profiles nearly unchanged, but increases the magnitude
of the imaginary vertical velocity and the near-surface extrema of the
streamwise response. Here $U_0$ is common to all cases, so the normalized
trends also describe the dimensional response. Since $\overline{U}(0)=0$,
the leading viscous length is $\delta_\nu=(\nu/kc)^{1/2}$ in every case,
with $\delta_\nu/\lambda=(2\pi\sqrt R)^{-1}\simeq5.0\times10^{-4}$. The
streamwise velocity consequently undergoes a rapid phase transition at
similar heights, changing from in phase with the elevation at the surface
to approximately antiphase above the viscous layer, while the vertical
velocity remains close to quadrature, $\phi_{\tilde w\eta}\simeq-\pi/2$.
In figure~\ref{fig:exp_sweep_delta} the WKBJ curves and the numerical solution are indistinguishable. Across these profiles and the companion sweep, the complex relative $L_2$ differences are below $0.36\%$ for vertical velocity and $0.45\%$ for streamwise velocity. These comparisons include the effect of truncating the numerical problem at $H$, whereas the WKBJ modes satisfy decay at infinity.

\begin{figure}[htbp]
\centering
\includegraphics[width=\textwidth]{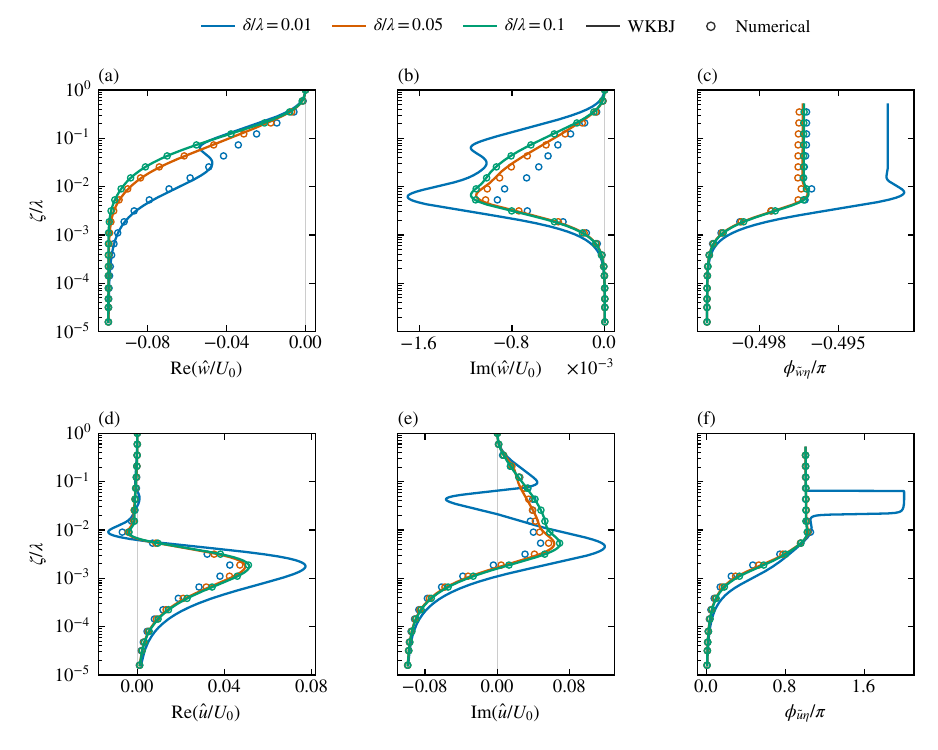}
\caption{As in figure~\ref{fig:exp_sweep_delta}, but for thin shear
layers, $\delta/\lambda=0.01$, $0.05$ and $0.10$ (blue, orange and green),
at $U_0/c=0.5$, $R=10^4$ and $ak=0.1$. Solid lines show the homogeneous
WKBJ approximation and open circles the full finite-difference solution.}
\label{fig:exp_thin_delta}
\end{figure}

To quantify the limits of the WKBJ solution, we test a case where the mean shear is confined to a very thin layer near the surface. Figure~\ref{fig:exp_thin_delta} shows
$\delta/\lambda=0.01$, $0.05$ and $0.10$ at $U_0/c=0.5$, $R=10^4$ and
$ak=0.1$. At $\delta/\lambda=0.10$, $\epsilon_Q\leq0.36$ and
$|\mathcal R_Q|\leq0.06$, and the WKBJ solution reproduces the full
numerical solution to within $0.9\%$ in vertical velocity and $1.5\%$ in
streamwise velocity. At $\delta/\lambda=0.05$, $\epsilon_Q$ reaches $1.0$,
and the differences increase to $3.7\%$ and $9.7\%$. At
$\delta/\lambda=0.01$, the shear layer spans only about six viscous
lengths, $\epsilon_Q$ and $|\mathcal R_Q|$ reach $6.1$ and $8.3$, and the
differences are larger, but with the correct qualitative trend. The weak components, namely the imaginary vertical
velocity and the streamwise response above the viscous layer, are the most
affected.
Despite this loss of accuracy, the construction captures the main parameter dependence and predicts velocity amplitudes of the correct order of magnitude.

The exponential profiles, therefore, support the two-branch description of
fast-wave airflow over smooth shear when the conditions of
\S\ref{sec:surface_matching} hold: wave kinematics set the surface motion,
and viscous matching supplies its phase adjustment. When the mean shear
is concentrated within a few viscous lengths of the surface, the outer
approximation breaks down. The logarithmic profile considered next tests
how the construction must be extended when the near-surface mean varies
too rapidly for the outer WKBJ approximation.

\subsection{Shifted logarithmic profile and near-wall long-wave construction}
\label{sec:log_longwave}

\subsubsection{Rapid near-surface variation}
Logarithmic mean velocity profiles are commonly used to describe the
approximately neutral atmospheric surface layer, and their strong
near-surface shear tests the assumptions of the WKBJ construction. Consider the shifted-logarithmic profile
\begin{equation}
\overline{U}(\zeta)=\frac{u_\ast}{\kappa}
\ln\left(1+\frac{\zeta}{z_0}\right),
\label{eq:longwave_log_profile}
\end{equation}
where $\kappa$ is the von K\'arm\'an constant and $z_0>0$ sets the
near-surface variation scale. The shift gives $\overline{U}(0)=0$ and finite surface
derivatives. The profile omits the viscous and buffer layers of a
turbulent boundary layer, and we use it to isolate the effect of rapid
mean-profile variation; the simulation-derived profiles of
\S\ref{sec:simulation_means} and \S\ref{sec:wrf_wall_modelled} supply resolved or modelled wall layers. Since the logarithm
increases without bound, the fast-wave condition $\overline{U}<c$ is imposed over a
finite interval containing the region dynamically coupled to the wave.

Using the signed wave-relative mean $V=\overline{U}-c$, the dimensional Rayleigh coefficient is $Q_{\mathrm d}=k^2+\overline{U}''/V$.
It is related to the nondimensional potential by
$Q_{\mathrm d}(\zeta)=k^2Q(k\zeta)$ and has units of inverse length squared. The diagnostics are invariant under this
change of scale: $\epsilon_Q=|Q_{\mathrm d}'|/Q_{\mathrm d}^{3/2}$ and
$\mathcal R_Q=5(Q_{\mathrm d}')^2/(16Q_{\mathrm d}^3)
-Q_{\mathrm d}''/(4Q_{\mathrm d}^2)$. For the shifted logarithmic profile,
$\overline{U}'=u_\ast/[\kappa(\zeta+z_0)]$ and
$\overline{U}''=-u_\ast/[\kappa(\zeta+z_0)^2]$. Consequently, $Q_{\mathrm d}>k^2$ wherever
$\overline{U}<c$, and the outer equation contains no critical level. The coefficient nevertheless varies rapidly near the surface when
$kz_0$ is small, so the slow-variation requirement of the outer WKBJ
approximation can fail there even though the equation is regular. This
limitation is inviscid: increasing the wave Reynolds number thins the
viscous layer but leaves the outer coefficient unchanged.

\subsubsection{Near-wall long-wave construction}

Near the wall we therefore represent the slow solution differently. Following the long-wave approach for the Rayleigh equation \citep{drazinhoward1962,bonfils2022}, we keep its dependence on the mean profile exact and expand it only in even powers of the horizontal wavenumber, $k$. The construction applies to any smooth, noncritical mean profile and is used for all the simulation-derived mean profiles below. Since the slow solution is the outer branch continued towards the wall, we retain the subscript $o$ and denote the dimensional outer velocity profile by $\widehat w_o(\zeta)$, which satisfies
\begin{equation}
\mathcal L_0\widehat w_o=k^2\widehat w_o,
\qquad
\mathcal L_0=\frac{\mathrm d^2}{\mathrm d\zeta^2}-\frac{\overline{U}''}{V}.
\label{eq:longwave_slow_operator}
\end{equation}
The expansion treats the explicit $k^2$ term as the perturbation,
with the dimensional mean profile, phase speed $c$ and inner
interval held fixed when constructing its coefficients.
The resulting series is evaluated at the physical wavenumber,
with $k$ and $c$ related by the deep-water dispersion relation.

When the explicit $k^2$ term is omitted from
\eqref{eq:longwave_slow_operator}, one solution is $y_1=V$,
since $V''=\overline U''$. A second
independent solution follows by reduction of order:
\begin{equation}
y_2(\zeta)=V(\zeta)I(\zeta),
\qquad
I(\zeta)=\int_0^\zeta\frac{\mathrm ds}{V^2(s)}.
\label{eq:longwave_zero_basis}
\end{equation}
These solutions have unit Wronskian, $y_1y_2'-y_1'y_2=1$. They retain the
full vertical dependence of the prescribed mean, including its rapid
near-surface variation. 
The unit Wronskian allows the forced reference equation to be expressed
through the integral operator
\begin{equation}
\mathcal K[f](\zeta)
=V(\zeta)\int_0^\zeta
V(s)\,[I(\zeta)-I(s)]f(s)\,\mathrm ds.
\label{eq:longwave_volterra_operator}
\end{equation}
Direct differentiation gives $\mathcal L_0\mathcal K[f]=f$, and both
$\mathcal K[f]$ and its first derivative vanish at the wall. For any combination $\mathcal{W}_0$  of
$y_1$ and $y_2$, the slow Rayleigh equation 
\eqref{eq:longwave_slow_operator} is therefore equivalent to the integral equation $\widehat w_o=\mathcal{W}_0+k^2\mathcal K[\widehat w_o]$, whose solution has the same wall value and slope as $\mathcal{W}_0$. Successive substitution yields the long-wave expansion
\begin{equation}
\widehat w_o
=\mathcal{W}_0+k^2\mathcal K[\mathcal{W}_0]+k^4\mathcal K^2[\mathcal{W}_0]+\cdots.
\label{eq:longwave_series}
\end{equation}
Writing $\mathcal{W}_{n+1}=\mathcal K[\mathcal{W}_{n}]$ for $n\geq0$, the expansion is
$\widehat w_o=\sum_{n=0}^{\infty}k^{2n}\mathcal{W}_n$. This is the series in $k^2$
introduced for the Rayleigh equation
\citep{drazinreid2004}.
Every correction has zero wall value and slope, so the two degrees of
freedom remain in the leading solution $\mathcal{W}_0$. The propagation kernel
depends on the mean velocity through $V$ and its integrals, without
requiring explicit evaluation of $\overline{U}''$. Mean derivatives remain necessary
for the outer WKBJ diagnostics and matching.
On a fixed finite interval where $|V|$ is bounded away from zero, the kernel is
bounded and the series \eqref{eq:longwave_series} converges for finite $k$. The
long-wave limit permits a short truncation, but a matching height farther
from the wall may require additional terms. For simplicity, the matched
comparisons below retain two terms, $\mathcal{W}_0+k^2\mathcal{W}_1$, with
$\mathcal{W}_1=\mathcal K[\mathcal{W}_0]$. The effect of retaining additional terms at the same matching height
is examined in Appendix~\ref{app:longwave_truncation}.

The inner solution is connected to the outer WKBJ mode at a height
$\zeta_m$ selected from the mean profile alone: we take $\zeta_m$ as the
lowest level above which $Q_{\mathrm d}>0$, $\epsilon_Q<0.3$ and
$|\mathcal R_Q|<0.3$ hold up to $H$, so that the outer WKBJ mode is valid
over the remaining propagation interval. Normalizing the decaying outer mode at the match
gives
\begin{equation}
B_o(\zeta)=
\left[\frac{Q_{\mathrm d}(\zeta)}{Q_{\mathrm d}(\zeta_m)}\right]^{-1/4}
\exp\left[-\int_{\zeta_m}^{\zeta}\sqrt{Q_{\mathrm d}(s)}\,\mathrm ds\right],
\label{eq:longwave_outer_mode}
\end{equation}
with dimensional logarithmic derivative $m_o=-\sqrt{Q_{\mathrm d}}-Q_{\mathrm d}'/(4Q_{\mathrm d})$. Let $Y_1$ and
$Y_2$ denote the long-wave solutions generated from $y_1$
and $y_2$ at the selected truncation order. The inner combination
$Y=a_1Y_1+a_2Y_2$ is selected by
\begin{equation}
\begin{pmatrix}
Y_1(\zeta_m)&Y_2(\zeta_m)\\
Y_1'(\zeta_m)&Y_2'(\zeta_m)
\end{pmatrix}
\begin{pmatrix}a_1\\a_2\end{pmatrix}
=\begin{pmatrix}1\\m_o(\zeta_m)\end{pmatrix}.
\label{eq:longwave_inner_matching}
\end{equation}
The resulting slow shape is continuous with $B_o$ in value and slope at
$\zeta_m$. Its overall complex
amplitude and that of the viscous mode are then determined
by imposing the two physical surface conditions on their sum.

The molecular viscous shape is $B_\nu(\zeta)=W_\nu(k\zeta)$, where
$W_\nu$ is the normalized decaying mode in \eqref{eq:viscous_modes};
hence $B_\nu(0)=1$. The complete approximation, which we refer to as the
long-wave WKBJ (LW--WKBJ) solution, is
\begin{equation}
\widehat w_c(\zeta)=
C_o\begin{cases}
Y(\zeta),&0\leq\zeta\leq\zeta_m,\\
B_o(\zeta),&\zeta\geq\zeta_m,
\end{cases}
+C_\nu B_\nu(\zeta).
\label{eq:longwave_composite}
\end{equation}
The wall logarithmic slope $m_\nu=B_\nu'(0)/B_\nu(0)=B_\nu'(0)$, like
$m_o$, has units of inverse length. The dimensional surface data
$\widehat w_s=-akc/2$ and $\widehat w_s'=ak[\overline U'(0)-kc]/2$, from
\eqref{eq:surface_fourier_amplitudes} and \eqref{eq:surface_conditions},
determine the amplitudes through
\begin{equation}
\begin{pmatrix}
Y(0)&1\\Y'(0)&m_\nu
\end{pmatrix}
\begin{pmatrix}C_o\\C_\nu\end{pmatrix}
=\begin{pmatrix}\widehat w_s\\\widehat w_s'\end{pmatrix}.
\label{eq:longwave_wall_matching}
\end{equation}
Surface matching can give the slow response a phase shift through its coupling to the viscous mode. This phase shift persists above the viscous layer.

\subsubsection{Shifted-logarithmic example}
\begin{figure}[htbp]
\centering
\includegraphics[width=\textwidth]{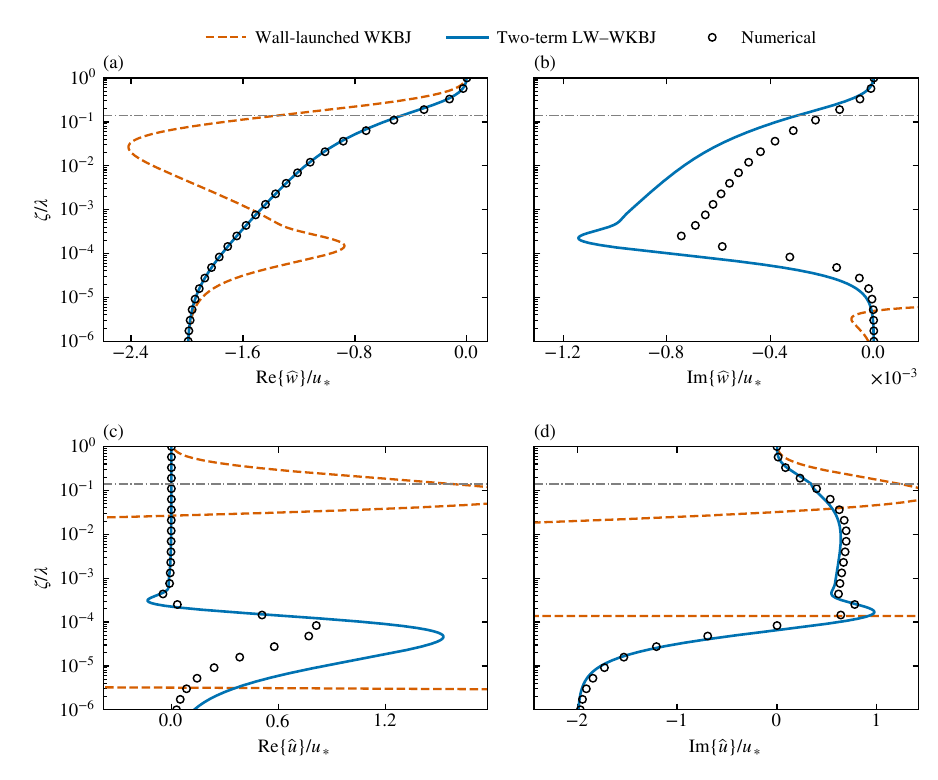}
\caption{Shifted-logarithmic example with $z_0=0.001\,\mathrm m$,
$u_\ast=0.25\,\mathrm{m\,s^{-1}}$ and $c/u_\ast=40$: (a,b) real and
imaginary vertical velocity and (c,d) real and imaginary streamwise
velocity, normalized by $u_\ast$. Orange dashed curves show the
wall-launched two-branch WKBJ solution, blue curves the two-term
LW--WKBJ and open circles the full numerical solution with mapped
forcing. The horizontal dash-dotted line marks the mean-selected
matching height. The wall-launched WKBJ solution leaves
the plotted range in panels (b--d).}
\label{fig:log_longwave_example}
\end{figure}

Figure~\ref{fig:log_longwave_example} illustrates the construction for
$z_0=0.001\,\mathrm m$, $u_\ast=0.25\,\mathrm{m\,s^{-1}}$, and
$c/u_\ast=40$. We use $\kappa=0.4$, $ak=0.1$, and molecular viscosity
$\nu=1.5\times10^{-5}\,\mathrm{m^2\,s^{-1}}$. The deep-water dispersion
relation gives $c=10\,\mathrm{m\,s^{-1}}$,
$k=0.0981\,\mathrm{m^{-1}}$, and $\lambda=64.05\,\mathrm m$.
The mean velocity
remains subcritical on the numerical interval $H=\lambda$,
and $R\simeq6.80\times10^6$. The matching criterion gives $\zeta_m/\lambda=0.141$, with
$\epsilon_Q(\zeta_m)=0.3$ and $|\mathcal R_Q(\zeta_m)|=0.195$.
The LW--WKBJ construction retains the two terms $\mathcal{W}_0+k^2\mathcal{W}_1$.
Figure~\ref{fig:log_longwave_example} shows that the wall-launched WKBJ
solution departs substantially from the numerical solution in every component, despite the large Reynolds number. This
failure follows from the rapid variation of the outer coefficient near the
wall: $\epsilon_Q$ and $|\mathcal R_Q|$ exceed unity for
$\zeta/\lambda\lesssim0.08$ and reach $7.75$ and $4.01$ at the surface.

However, replacing the near-wall WKBJ outer mode by the long-wave construction removes
this failure. The relative $L_2$ discrepancy in $\Real\{\widehat w\}$
over $0\leq\zeta\leq H$ decreases from $171\%$ to $5.0\%$, and the
dominant streamwise component, $\Imag\{\widehat u\}$, agrees with the
numerical solution to within $10\%$. The LW--WKBJ also reproduces the
structure and sign of the much smaller components $\Imag\{\widehat w\}$,
of order $10^{-3}u_\ast$, and $\Real\{\widehat u\}$, which is confined to
the viscous layer, although it overestimates their peaks there. 
The accuracy of the two-term truncation, and the effect of retaining higher-order terms, are assessed in Appendix~\ref{app:longwave_truncation}.

The shifted logarithmic profile therefore shows that rapid near-surface
variation, which is typical of turbulent mean profiles over waves, can
invalidate a WKBJ solution launched at the surface even when the Rayleigh
equation has no critical level. The LW--WKBJ construction
bridges this region while retaining WKBJ propagation farther from the
wall, so the construction remains analytical and accurate for such
profiles. The simulation-derived velocity profiles 
considered below
and the wall-modelled LES described in \S\ref{sec:wrf_wall_modelled} use the same LW--WKBJ construction.

\subsection{Simulation-derived mean profiles}
\label{sec:simulation_means}
\subsubsection{Wall-resolved LES}
\label{sec:cao_wrles}
The wall-resolved simulations of \citetalias{caoshen2021} provide a mean
profile with a resolved viscous sublayer and buffer layer. They showed that numerical solutions of the
wave-following Orr--Sommerfeld equation reproduce the LES
wave-coherent fields at high wave ages, so the LW--WKBJ is directly compared
 with the LES data. We consider
their fast wave case with $c/u_\tau=54.3$, $c/U_0=1.2$,
$U_0\lambda/\nu=30000$ and $ak=0.15$. Here $u_\tau$ is the air-side
friction velocity supplied with these simulations, corresponding to
$u_\ast$ in the preceding sections, and $U_0$ is the reference velocity
of the simulation. The wave Reynolds number is $R=c/(k\nu)\simeq5730$.
The digitized mean velocity is represented by the constrained polynomial \eqref{eq:app_mean_fit} of Appendix~\ref{app:mean_fits}, written in wall units $\overline U^+=\overline U/u_\tau$ and $z^+=\zeta u_\tau/\nu$. Its form enforces $\overline U(0)=0$ and the viscous wall slope $\overline U'(0)=u_\tau^2/\nu$, so that $\overline U^+=z^+$ near the wall, and all derivatives are obtained analytically.

The matched LW--WKBJ retains the two-term truncation $\mathcal{W}_0+k^2\mathcal{W}_1$ of
\eqref{eq:longwave_series}, adds the molecular viscous mode, and is
connected to the outer WKBJ branch as in
\eqref{eq:longwave_inner_matching}. Applied at the digitized levels, the
matching criterion of \S\ref{sec:log_longwave} gives $z_m^+=116.7$ or $k\zeta_m=1.11$, where $\epsilon_Q=0.25$ and
$|\mathcal R_Q|=0.15$. Below this height the outer approximation
deteriorates rapidly through the buffer layer; $\epsilon_Q$ reaches
$0.73$ at $z^+=75$ and exceeds $2$ between $z^+=10$ and $30$. 
The streamwise velocity includes the mapped-continuity term in
\eqref{eq:WKBJ_reconstruct_u}, with $G(\zeta)=\zeta/H-1$ and $H=\lambda$,
and velocities are normalized by $u_\tau$. The analytic mean is
continued to $H$, and the pressure
reconstruction is described in \S\ref{sec:cao_pressure}.

\begin{figure}[htbp]
\centering
\includegraphics[width=\textwidth]{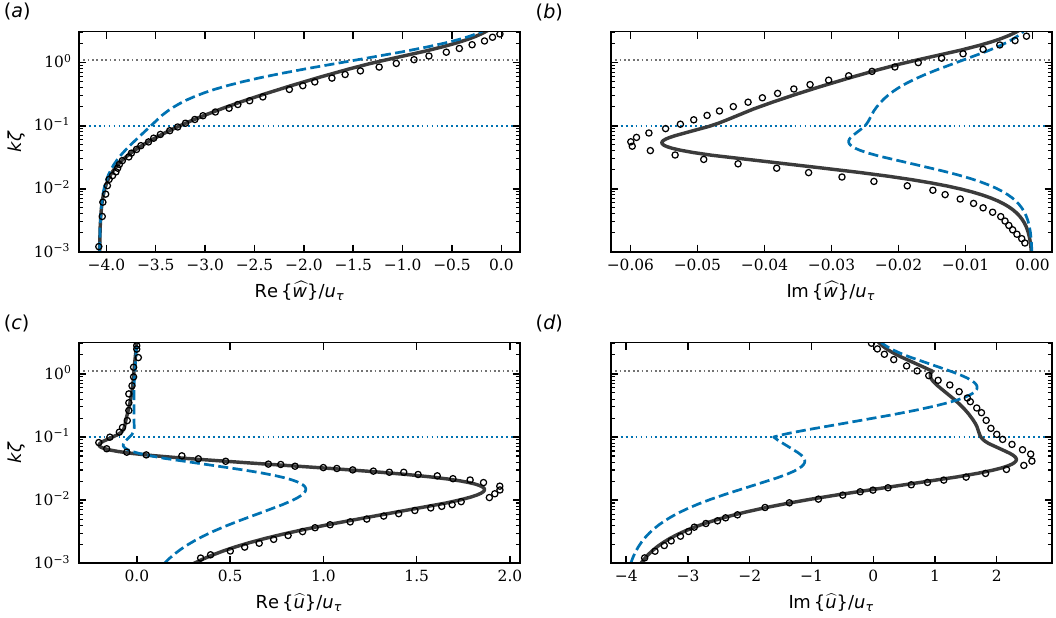}
\caption{Wave-coherent velocity for the $c/u_\tau=54.3$
case of \citetalias{caoshen2021}: (a,b) real and imaginary vertical
velocity and (c,d) real and imaginary streamwise velocity. Open circles denote digitized LES results and solid curves
the LW--WKBJ. The grey dotted line marks the
matching height $z_m^+=116.7$ ($k\zeta_m=1.11$), selected from the
mean-profile diagnostics. For illustration, the dashed blue curves show the
LW--WKBJ matched instead at $k\zeta=0.1$ ($z^+\approx10.6$, blue dotted
line), inside the buffer layer, where $\epsilon_Q=3.1$ and
$|\mathcal R_Q|=0.50$.}
\label{fig:cao_shen_harmonics}
\end{figure}

Figure~\ref{fig:cao_shen_harmonics} compares the LW--WKBJ 
 with the real
and imaginary parts of the LES vertical and streamwise velocity.
Evaluated at the digitized LES points, the LW--WKBJ reproduces the LES
velocity with discrepancies of
$4.3\%$ in $\Real\{\widehat w\}$ and $4.8\%$ and $9.0\%$ in the real and
imaginary parts of $\widehat u$. It captures the near-surface extrema of
$\widehat u$ and the decay of $\widehat w$. The largest departures in these
components occur in $\Imag\{\widehat u\}$ below the match, where the
LW--WKBJ underestimates the peak by about $10\%$, and in
$\Real\{\widehat w\}$ above it, where it overestimates the magnitude. Both
arise from truncating the long-wave series at the relatively large
matching height $k\zeta_m\approx1.1$; retaining a third term reduces the
three discrepancies to $2.3\%$, $2.8\%$ and $3.4\%$
(Appendix~\ref{app:longwave_truncation}). The phase-shifted vertical
velocity $\Imag\{\widehat w\}$, which peaks at about $1.5\%$ of the
surface value of $|\Real\{\widehat w\}|$, differs from the LES by $13\%$.
Matching instead inside the buffer layer (dashed blue curves in figure~\ref{fig:cao_shen_harmonics}) changes $\Real\{\widehat w\}$ only slightly but degrades the streamwise velocity: the near-surface peak of $\Real\{\widehat u\}$ is underestimated, and $\Imag\{\widehat u\}$ is distorted around the matching height. The order of magnitude and the qualitative structure are nevertheless retained.

In summary, the LW--WKBJ construction accurately reproduces the wave-induced velocity structure of the wall-resolved LES, capturing both the near-surface extrema and the subsequent decay with height. The corresponding pressure response is examined in \S\ref{sec:cao_pressure}.

\subsubsection{Coupled air--sea DNS}
\label{sec:zhang_swell}
The coupled air--sea DNS of \citetalias{zhang2025windwave} provides a second
mean profile, at a lower Reynolds number where the wind and the wave are fully coupled. We consider their swell case
with $c/u_\ast=27.7$, $ak=0.25$ and $Re_\tau=u_\ast\lambda/\nu=180$, for
which $R=c/(k\nu)\approx790$. Their dataset consists of the plane-averaged
air-side velocity and stresses but not the wave-coherent fields. The LW--WKBJ is therefore compared with
the numerical solution of the wave-following Orr--Sommerfeld equation,
both computed from the same mean profile and wave parameters.

The DNS mean is a plane average at fixed height $Z$ above the mean water
level. Far from the surface, the mapped and physical heights coincide, and
we identify $\zeta=Z$. Close to the surface, however, the plane average at fixed $Z$ differs from
the mean in the wave-following coordinate $\zeta$. The two agree to first
order in the surface displacement but differ at second order, through the
mean-profile curvature and the correlation between the wave-coherent shear
and the displacement. These terms are large within a few wave amplitudes of
the interface, where the viscous and buffer layers lie at this Reynolds
number. The data from \citetalias{zhang2025windwave} show that the plane average is not monotonic between the crest and about $1.3$
amplitudes above it. We therefore discard all samples within two wave amplitudes of the crest, $Z<Z_{\rm crest}+2a\approx0.125\lambda$, where $Z_{\rm crest}=[ak+(ak)^2/2+3(ak)^3/8]/k$ is the crest height of the initial third-order Stokes wave. The remaining 162 samples, which span
$0.132\leq\zeta/\lambda\leq1.25$, are fit to the constrained polynomial form in Appendix~\ref{app:mean_fits}.
The retained samples begin at $z^+\approx24$, so the viscous sublayer is not constrained by data. We therefore also impose zero wall curvature, $\mathrm d^2\overline U^+/\mathrm dz^{+2}(0)=0$.
The LW--WKBJ again retains $\mathcal{W}_0+k^2\mathcal{W}_1$ and the molecular viscous mode,
with the surface conditions given by \eqref{eq:surface_conditions}.
Moving the cut-off between $1.5a$ and $3a$ changes the matching
height from $k\zeta_m=0.92$ to $0.96$ and the form-stress coefficient by
less than $2\%$.
The numerical solution retains the geometric forcing
in \eqref{eq:OS_dimensional} and clusters points near the surface to maintain accuracy.
\begin{figure}[htbp]
\centering
\includegraphics[width=0.9\textwidth]{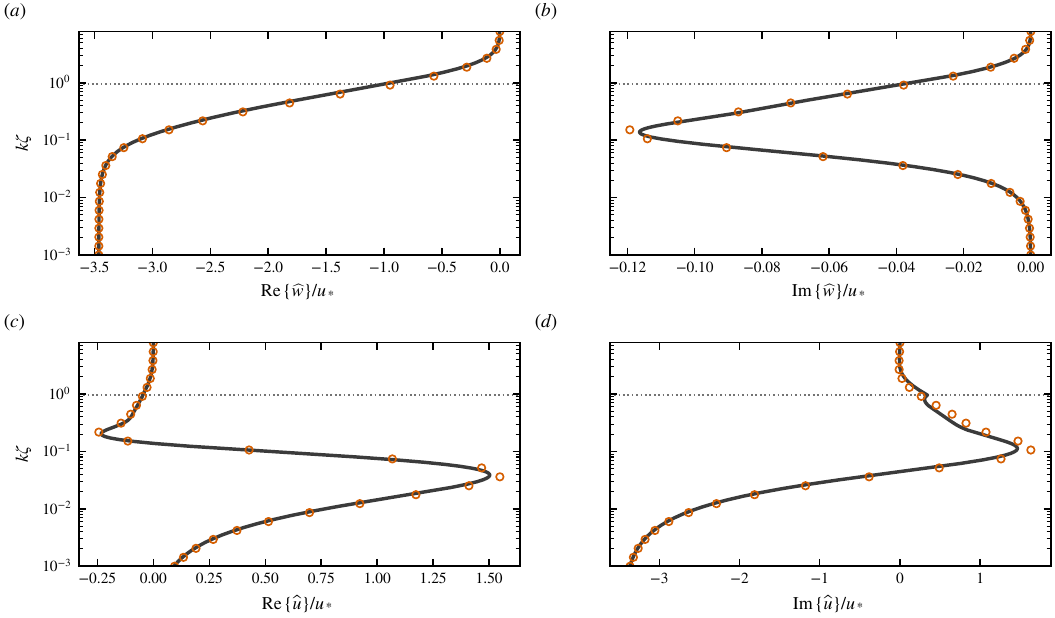}
\caption{Wave-coherent velocity for the coupled-swell mean of
\citetalias{zhang2025windwave}: (a,b) real and imaginary vertical velocity
and (c,d) real and imaginary streamwise velocity. Solid curves show the LW--WKBJ and open circles the numerical solution
using the simulation-derived mean velocity. The dotted line marks the
matching height $z_m^+\approx27.5$. Heights are measured from the mean
water level.}
\label{fig:zhang_coupled_swell_velocity}
\end{figure}
Figure~\ref{fig:zhang_coupled_swell_velocity} compares the LW--WKBJ with the
numerical solution of the viscous Orr--Sommerfeld equation using the DNS mean profile and shows close agreement in
all four components. The
dominant vertical velocity $\Real\{\widehat w\}$ equals the surface value
$-akc/2$ at the interface and decays smoothly, falling to half of it at
$k\zeta\approx0.5$ (panel a). The phase-shifted component
$\Imag\{\widehat w\}$ is small, peaking at about $3.5\%$ of the surface
value near $k\zeta\approx0.14$ (panel b). The streamwise velocity has a
clear viscous layer at the surface (panels c,d). No slip ties
$\Imag\{\widehat u\}$ to the orbital velocity of the surface, $-akc/2$, at
$\zeta=0$. Within $k\zeta\approx0.04$, about the Stokes-layer thickness at
$R\approx790$, $\Imag\{\widehat u\}$ changes sign and reaches the inviscid
response, which opposes the orbital motion and peaks near $k\zeta\approx0.1$
before decaying. The phase-shifted component $\Real\{\widehat u\}$ is
generated in the same layer: it vanishes at the surface, peaks at about
$45\%$ of the surface orbital velocity near $k\zeta\approx0.04$ and changes
sign at $k\zeta\approx0.14$. 
The LW--WKBJ reproduces all of these features and shows good agreement in all four velocity components, including the small phase-shifted ones.

This section has shown that the WKBJ and LW--WKBJ constructions reproduce
the wave-coherent velocity for analytical mean profiles and for mean profiles
from wall-resolved simulations, in which the viscous sublayer and the
near-surface shear are resolved. Simulations of the atmospheric boundary
layer at high Reynolds number rarely resolve this region, because the cost
is prohibitive, and the surface stress is supplied by a wall model.
Section~\ref{sec:wrf_wall_modelled} extends the construction to wall-modelled LES
by adding closures for the unresolved near-wall mean and viscous layer.


\section{Wall-modelled large-eddy simulations}
\label{sec:wrf_wall_modelled}
To test the approach when the viscous and buffer layers are unresolved, we conduct wall-modelled, phase-resolved LES using a moving-wave implementation of the Weather Research and
Forecasting (WRF) model developed by \citet{zhu2023moving}. We consider two swell cases
with $c/u_\tau=54.5$ and $81.0$ and $ak=0.10$.  the simulations of
\citetalias{caoshen2021} and \citetalias{zhang2025windwave}, the WRF simulations use an equilibrium wall model to supply the surface stress, and the first resolved level lies above the
molecular wall layer. The LW--WKBJ, therefore, 
specifies the effective viscosity for the viscous mode solution. Section~\ref{sec:wrf_setup}
describes the simulations, \S\ref{sec:wrf_closures} the viscosity closure, and
\S\ref{sec:wrf_velocity} the wave-coherent velocity.

\subsection{Numerical method}
\label{sec:wrf_setup}
The lower boundary of the simulation follows a prescribed
propagating sinusoid using a terrain-following approach. Its levels are displaced by $h=(1-\zeta/H)\eta$, the mapping of
\S\ref{sec:model} with $G=\zeta/H-1$ and $H=100\,\mathrm m$, so the
harmonics on the model levels correspond directly to the wave-following
fields of the theory, including the mean-shear term in
\eqref{eq:WKBJ_reconstruct_u}.
The configuration follows the turbulent moving-wave case of
\citet{zhu2023moving}, and the simulation parameters are provided in table~\ref{tab:wrf_les}.
The domain is periodic in the streamwise and spanwise ($x$ and $y$) directions with six
wavelengths in the streamwise direction. The flow is neutral, dry and
non-rotating, and is driven by a constant streamwise pressure gradient.
Subgrid stresses are represented by a Smagorinsky closure \citep{smagorinsky1963}. At the moving
surface, the stress is given by a quadratic drag law applied to the
velocity relative to the surface at the lowest grid level, with a drag
coefficient chosen so that this velocity satisfies the logarithmic law
for the roughness length $z_0$ \citep{zhu2023moving}. The lowest mass
level therefore lies within a modelled wall layer.

WRF applies the imposed pressure gradient divided by the local air
density, so the friction velocity is
\begin{equation}
u_\tau=\left(\left|\frac{\mathrm dp}{\mathrm dx}\right|
\int_0^H\frac{\mathrm d\zeta}{\rho}\right)^{1/2},
\label{eq:wrf_utau}
\end{equation}
which we evaluate from the simulated mean density and use throughout. The
corresponding effective density, $H/\int_0^H\rho^{-1}\,\mathrm d\zeta$, is
$1.168$\,kg\,m$^{-3}$ in both cases. The mean surface stress, including the
form drag, gives a friction velocity $u_\ast$ within $1$--$2\,\%$ of
$u_\tau$, consistent with the small drift of the surface stress over the
averaging window.

\begin{table}
\centering
\begin{tabular}{lcc}
\hline
 & $c/u_\tau=54.5$ & $c/u_\tau=81.0$ \\
\hline
Domain, $L_x\times L_y\times H$ & \multicolumn{2}{c}{$300\times150\times100$\,m ($6\lambda\times3\lambda\times2\lambda$)} \\
Grid points & \multicolumn{2}{c}{$120\times60\times80$} \\
Grid spacing & \multicolumn{2}{c}{$\Delta x=\Delta y=2.5$\,m; $\Delta z=0.51$\,m at the surface, stretched by $1.02$} \\
Time step & \multicolumn{2}{c}{$0.01$\,s, four acoustic substeps} \\
Wave & \multicolumn{2}{c}{$\lambda=50$\,m, $a=0.8$\,m, $ak=0.10$, $c=8.84$\,m\,s$^{-1}$} \\
Subgrid model; roughness length & \multicolumn{2}{c}{Smagorinsky, $C_s=0.18$; $z_0=2\times10^{-4}$\,m} \\
Pressure gradient, $\mathrm dp/\mathrm dx$ (Pa\,m$^{-1}$) & $-3.07\times10^{-4}$ & $-1.39\times10^{-4}$ \\
$u_\tau$ from \eqref{eq:wrf_utau} (m\,s$^{-1}$) & $0.1622$ & $0.1090$ \\
$c/u_\tau$ & $54.5$ & $81.0$ \\
$R=c/(k\nu)$ & \multicolumn{2}{c}{$4.69\times10^{6}$} \\
$R_{\rm eff}=c/[k\nu_{\rm eff}(0)]$ & $2.93\times10^{4}$ & $4.14\times10^{4}$ \\
\hline
\end{tabular}
\caption{Configuration of the wall-modelled WRF simulations. The friction
velocity $u_\tau$, used for normalization, follows from the imposed
pressure gradient and the simulated density through
\eqref{eq:wrf_utau}; $R$ uses the molecular viscosity
$\nu=1.5\times10^{-5}$\,m$^2$\,s$^{-1}$, and $R_{\rm eff}$ the effective
viscosity \eqref{eq:nu_eff} at the surface.}
\label{tab:wrf_les}
\end{table}

Statistics are accumulated over a statistically stationary period of each
run. Fields are conditionally averaged on the
native model levels in 72 bins of the wave phase
$\theta=k\xi=kx-\omega t$, with $\omega=kc$. A field $f$ is decomposed into
its average over $x$, $y$ and $t$ on each level, the phase-averaged
wave-coherent component $\widetilde f(\theta,\zeta)$ and the turbulent
remainder $f'$. The amplitudes $\widehat f$ are the first Fourier
coefficients of $\widetilde f$ under \eqref{eq:fourier_convention},
weighted by the number of samples in each phase bin.

\begin{figure}
\centering
\includegraphics[width=\textwidth]{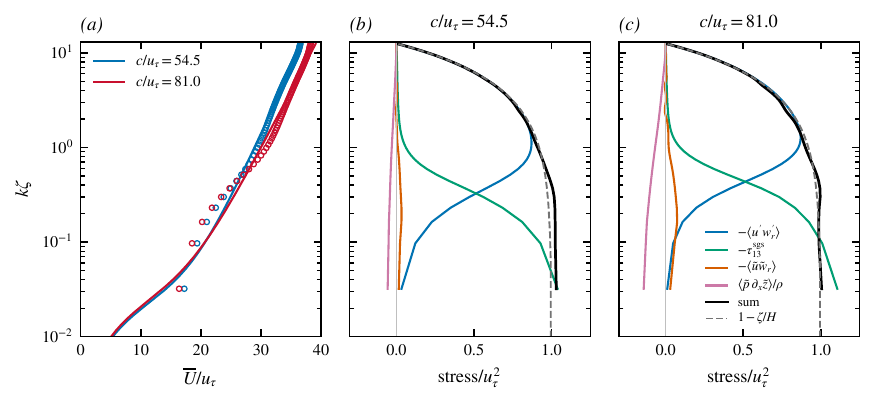}
\caption{Mean flow in the WRF simulations. (a) Mean streamwise velocity
at the native levels (circles) and the Spalding completions of
\S\ref{sec:wrf_closures} (curves). (b,c) Mean downward flux of streamwise momentum across the moving
model levels: resolved turbulent and wave-coherent fluxes based on the
velocity relative to the level, $w_r=w-(u-c)\,\partial_x Z$; subgrid
stress; and pressure stress on the sloping levels,
$\langle\tilde p\,\partial_x\tilde Z\rangle/\rho$. The black curve is their
sum and the dashed line is $1-\zeta/H$. }
\label{fig:wrf_les_mean_stress}
\end{figure}

Figure~\ref{fig:wrf_les_mean_stress} shows the mean flow. The mean velocity (panel a) remains below the phase speed throughout the domain, so neither case has a critical level. Panels (b,c) show the mean momentum flux across the moving model levels, evaluated with the vertical velocity relative to the level, $w_r=w-(u-c)\,\partial_x Z$, where $Z$ is the physical height of the level. Near the surface the stress is carried almost entirely by the subgrid closure. The wave-coherent flux and the pressure stress remain small, reaching about $0.02$ and $0.07u_\tau^2$, and $-0.05$ and $-0.14u_\tau^2$, for the two cases; at the lowest level the pressure stress agrees with the WRF form-drag diagnostic to within $2\,\%$.

\subsection{Near-wall closure}
\label{sec:wrf_closures}
The  mean velocity is represented by the implicit law of \citet{spalding1961}. With $f=\overline U/u_\tau$, the law reads
\begin{equation}
\frac{\zeta S_0}{u_\tau}
=f+\mathrm e^{-\kappa B_{\mathrm{Sp}}}
\left[\mathrm e^{\kappa f}-1-\kappa f
-\frac{(\kappa f)^2}{2}-\frac{(\kappa f)^3}{6}\right].
\label{eq:wrf_spalding}
\end{equation}
The wall slope is defined as $S_0=\overline U(\zeta_1)/\zeta_1$, the secant between the surface and the lowest model level, $\zeta_1=0.26$\,m ($k\zeta_1=0.032$), at which the wall model applies its drag law, and $z^+=\zeta S_0/u_\tau$. The completed profile satisfies $\overline U(0)=0$. Only $\kappa$ and $B_{\mathrm{Sp}}$ are fitted, using the mean velocities at all model levels (table~\ref{tab:wrf_mean_fits}). The fits reproduce the WRF mean to within $1.6\%$ and $2.4\%$
(figure~\ref{fig:wrf_les_mean_stress}a) and supply all derivatives analytically.
The viscous mode cannot use the molecular viscosity, because the
near-surface stress is carried almost entirely by the wall model and the
subgrid closure (\S\ref{sec:wrf_setup}). Rather than extracting an eddy
viscosity from the subgrid model, we replace $\nu$ in the viscous mode by
the constant-stress effective viscosity
\begin{equation}
\nu_{\rm eff}(\zeta)=\max\!\left(\nu,\frac{u_\tau^2}{\overline U'(\zeta)}\right),
\label{eq:nu_eff}
\end{equation}
with the molecular value $\nu=1.5\times10^{-5}\,\mathrm{m^2\,s^{-1}}$ acting as a
floor. At the wall $\nu_{\rm eff}=u_\tau^2/S_0=2.4\times10^{-3}$ and
$1.7\times10^{-3}\,\mathrm{m^2\,s^{-1}}$, so the effective Stokes layer, of
thickness $(\nu_{\rm eff}/kc)^{1/2}\approx0.05$ and $0.04$\,m, lies below the
lowest model level. In \eqref{eq:viscous_modes}, $R^{1/2}\int_0^z\Delta^{1/2}\,\mathrm ds$
is replaced by $\int_0^\zeta[k(c-\overline U)/\nu_{\rm eff}]^{1/2}\,\mathrm d\zeta'$,
and the viscous stress of this mode, with $\nu_{\rm eff}(\zeta)$, is retained
in the pressure reconstruction since it supplies the phase shift that produces
the form stress (\S\ref{sec:surface_matching}). The matching criterion of
\S\ref{sec:log_longwave} gives $k\zeta_m=0.87$ and $0.74$ for
$c/u_\tau=54.5$ and $81.0$, respectively.

\subsection{Wave-coherent velocity}
\label{sec:wrf_velocity}
\begin{figure}[tbp]
\centering
\includegraphics[width=0.8\textwidth]{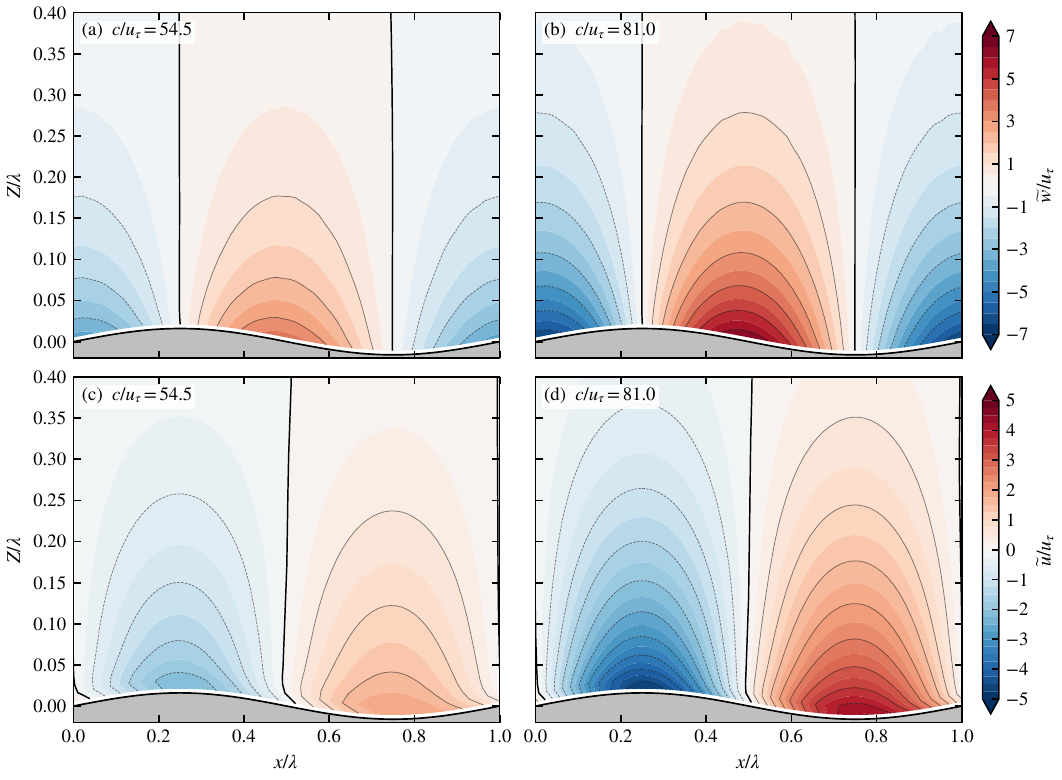}
\caption{Phase-averaged wave-coherent velocity in the WRF LES at
(a,c) $c/u_\tau=54.5$ and (b,d) $c/u_\tau=81.0$, over one wavelength:
(a,b) vertical velocity $\widetilde w/u_\tau$ and (c,d) streamwise velocity
$\widetilde u/u_\tau$. The
wave propagates in the positive $x$ direction, with the crest at
$x/\lambda=0.25$. The fields are plotted on the model levels at their
phase-averaged heights; thin contours are spaced by $u_\tau$ in (a,b) and
$0.5u_\tau$ in (c,d) and dashed where negative, and the thick contour marks
zero. The grey region is the water.}
\label{fig:wrf_w_contours}
\end{figure}
Figure~\ref{fig:wrf_w_contours} shows the phase-averaged wave-coherent
velocity over one wavelength. In both cases $\widetilde w$ is upward ahead
of the crest and downward behind it, nearly in quadrature with the surface
elevation, and its zero contours stay close to vertical over the height
shown. Its magnitude is set by the surface value $akc$ and
decays over approximately $0.2$--$0.3\lambda$. The streamwise velocity
$\widetilde u$ is negative over the crest and positive over the trough,
opposite to the orbital velocity of the surface. Its zero contours bend
just above the surface: at the lowest level $\widetilde u$ is shifted in
phase by about $11^\circ$ and $5^\circ$ from the response above, where the
shift is below $2^\circ$, as the wall model turns the air toward the
orbital velocity of the surface.
\begin{figure}[tbp]
\centering
\includegraphics[width=\textwidth]{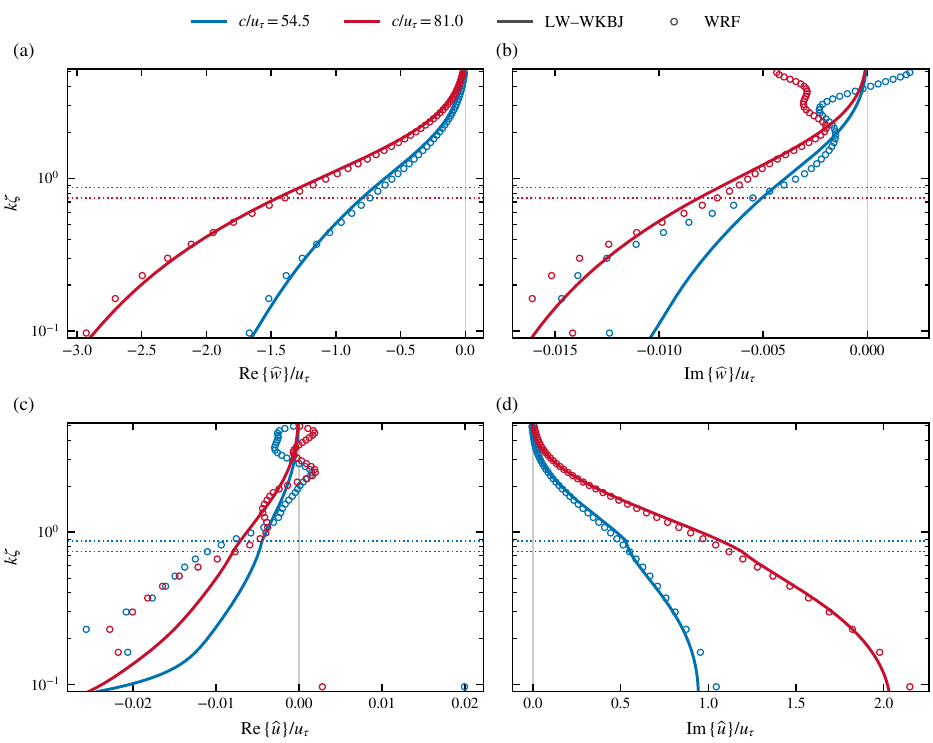}
\caption{Wave-coherent velocity for the wall-modelled WRF LES at
$c/u_\tau=54.5$ (blue) and $81.0$ (red): (a) real and (b) imaginary vertical
velocity, and (c) real and (d) imaginary streamwise velocity. Curves show
the two-term LW--WKBJ solution with the
effective-viscosity viscous mode, and open circles the WRF harmonics on the
resolved levels; dotted lines mark the matching heights.}
\label{fig:wrf_velocity}
\end{figure}
Figure~\ref{fig:wrf_velocity} compares the LW--WKBJ with the WRF
harmonics over the resolved levels, $0.097\leq k\zeta\leq5$. The dominant
components agree closely, with the complex relative $L_2$ discrepancies below approximately $7\%$. As in the \citetalias{caoshen2021} case, retaining a third term in the LW--WKBJ expansion reduces the
vertical-velocity discrepancies further (Appendix~\ref{app:longwave_truncation}). The
phase-shifted components $\Imag\{\widehat w\}$ and $\Real\{\widehat u\}$
are only about $1\%$ and $2\%$ of the dominant components. The LW--WKBJ gives their sign and order of magnitude, except for $\Real\{\widehat u\}$ at the lowest model level, $k\zeta = 0.097$.

Using an effective-viscosity viscous mode, the
LW--WKBJ therefore reproduces the dominant wave-coherent velocity of the
wall-modelled simulations.

\section{Wave-induced pressure and form stress}
\label{sec:pressure_traction_coefficients}
The correlation between the wave-coherent surface pressure and
the surface slope determines the horizontal form stress exerted
by the air on the water \citep{sullivan2000,wu2022}. This stress is a force per unit horizontal
area. For $\eta=a\sin(k\xi)$ the surface slope is $\eta_\xi\equiv\partial\eta/\partial\xi=ak\cos(k\xi)$, which is dimensionless. Averaging the product of the wave-coherent surface pressure and the slope over a wavelength, denoted $\langle\cdot\rangle_\xi$, and using \eqref{eq:fourier_convention} gives
\begin{equation}
F_p=\left\langle\widetilde p_s\eta_\xi\right\rangle_\xi
=ak\,\operatorname{Re}\{\widehat p_s\}
=\frac12\rho_a u_\ast^2(ak)^2\beta,
\label{eq:form_stress}
\end{equation}
where $u_\ast$ is the friction velocity used for each case. Thus
$\beta=2\operatorname{Re}\{\widehat p_s\}/(\rho_a u_\ast^2ak)$. This is
the usual normalization of the form stress
\citep{townsend1972,plant1982}: for a free gravity wave forced by
the pressure alone, the fractional energy gain per radian of phase is
$\beta(\rho_a/\rho_w)(u_\ast/c)^2$. A positive
$\beta$ denotes a pressure force on the water in the direction of wave
propagation; a negative $\beta$ denotes a force against it, so that the
air gains momentum in the propagation direction. 

For the fast-wave cases considered in \S\ref{sec:simulation_means} and \S\ref{sec:wrf_wall_modelled}, the slope-correlated component $\Real\{\widehat p_s\}$ is only $1$--$4\%$ of the surface-pressure amplitude. Predicting the form stress therefore requires accurately resolving a small phase-shifted component, providing a more stringent test of the solution than agreement in the overall pressure amplitude. To express the form stress using the wave speed as the velocity scale, we introduce
\begin{equation*}
\overline{\beta}=(u_\ast/c)^2\beta
=\frac{2F_p}{\rho_a c^2(ak)^2}.
\end{equation*}
This normalization is convenient for stating the leading-order dependence on the wave Reynolds number. Section~\ref{sec:leading_form_stress} derives the leading-order
form stress, and \S\ref{sec:exponential_form_stress} compares
the WKBJ and numerical results for the exponential profile.
Section~\ref{sec:cao_pressure} examines the pressure profiles
for \citetalias{caoshen2021}, \citetalias{zhang2025windwave}
and the wall-modelled WRF simulations.
Section~\ref{sec:simulation_beta} then compares the corresponding
surface form-stress coefficients.
\subsection{Leading-order form stress}
\label{sec:leading_form_stress}
The asymptotic solution gives an explicit expression for the leading-order
form stress. The $O(R^{-1/2})$ viscous correction in
\eqref{eq:mode_amplitude_scaling} supplies the phase shift that produces
a pressure component correlated with the surface slope.
 The resulting expression depends on two quantities.
\begin{align}
\mathcal M
&\equiv \overline u'(0)-1+W_o'(0),
\label{eq:matching_defect}\\
\mathcal I[\overline u]
&\equiv \int_0^\infty \Delta(z)W_o(z)\,\mathrm dz,
\label{eq:outer_profile_integral}
\end{align}
where $W_o$ is the normalized decaying outer WKBJ mode.
The matching term $\mathcal M$ measures the adjustment supplied by
the Stokes mode, while $\mathcal I$ relates the outer velocity
response to the surface pressure.

Under the fixed-profile, large-$R$ assumptions of
\S\ref{sec:surface_matching},
$W_\nu'(0)=\mathrm e^{3\mathrm i\pi/4}\sqrt R+O(1)$.
Using this in \eqref{eq:matched_amplitudes} gives
$\operatorname{Im}A_o=ak\mathcal M/(2\sqrt{2R})+O(akR^{-1})$.
The small amplitude and thickness of the Stokes mode make its direct
contribution to $\widehat p_s/(\rho_a c^2)$ only $O(akR^{-1})$,
as are the explicit viscous and geometric-forcing contributions. Equation~\eqref{eq:WKBJ_reconstruct_p}
therefore gives
$\widehat p_s/(\rho_a c^2)
=-\mathrm iA_o\mathcal I+O(akR^{-1})$.
Since $\mathcal I$ is real, the definition of $\overline\beta$ yields
\begin{equation}
\overline\beta
=\frac{\mathcal M\,\mathcal I[\overline u]}{\sqrt{2R}}
+O(R^{-1}).
\label{eq:leading_transfer_coefficients}
\end{equation}
For the exact Rayleigh mode, integrating
$(\Delta W_o'-\Delta'W_o)'=\Delta W_o$ from the surface to infinity,
using $\Delta(0)=W_o(0)=1$, gives
\begin{equation}
\mathcal M=-\left(1+\mathcal I[\overline u]\right).
\label{eq:defect_identity}
\end{equation}
This identity holds to the accuracy of the outer WKBJ approximation,
so
\begin{equation}
\overline\beta
=-\frac{\mathcal I[\overline u]\bigl(1+\mathcal I[\overline u]\bigr)}
{\sqrt{2R}}
+O(R^{-1}),
\label{eq:reduced_transfer_coefficients}
\end{equation}
where the remainder denotes finite-$R$ corrections, separate from
the outer WKBJ error.

The leading-order result links the prescribed mean wind directly to
the pressure form stress. The viscous surface layer supplies the phase
shift and sets the $R^{-1/2}$ dependence, while the outer response
throughout the overlying air determines the coefficient through
$\mathcal I$. For the present fast-wave regime, $\mathcal I>0$, so the
pressure force transfers momentum from the waves to the air without
requiring a critical layer. Since \eqref{eq:defect_identity} holds
exactly for the decaying Rayleigh solution,
\eqref{eq:reduced_transfer_coefficients} also applies when $W_o$ is the
exact or long-wave Rayleigh solution; for the wall-modelled cases $R$ is
replaced by $R_{\rm eff}$ (\S\ref{sec:wrf_closures}).

\subsection{Form stress for the exponential profile}
\label{sec:exponential_form_stress}
\begin{figure}[!htbp]
\centering
\includegraphics[width=\textwidth]{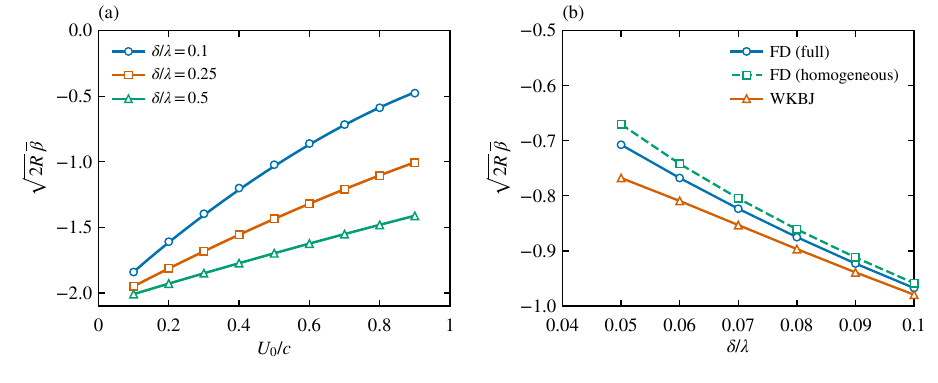}
\caption{Rescaled pressure form stress $\sqrt{2R}\,\overline\beta$ for the
exponential mean profile at $ak=0.1$, where
$\overline\beta=2F_p/[\rho_a c^2(ak)^2]$. (a) $R=10^3$, as a function of
$U_0/c$, for $\delta/\lambda=0.1$, $0.25$ and $0.5$. Solid curves show the
homogeneous two-mode WKBJ approximation evaluated with the closed-form
exponential phase integrals; open symbols show the full finite-difference
solution, including geometric forcing. (b) Thin shear layers at
$U_0/c=0.5$ and $R=10^4$, as a function of $\delta/\lambda$, from the full
and homogeneous finite-difference solutions and the homogeneous WKBJ
approximation. }
\label{fig:exponential_form_stress_comparison}
\end{figure}

Figure~\ref{fig:exponential_form_stress_comparison}(a) compares the WKBJ
and numerical solution values of $\sqrt{2R}\,\overline\beta$ for the
exponential profile at $R=10^3$ and $ak=0.1$. The wind-speed ratio
$U_0/c$ varies from $0.1$ to $0.9$ at fixed wave speed, viscosity and
steepness, for three values of $\delta/\lambda$. The factor $\sqrt{2R}$
removes the leading $R^{-1/2}$ dependence in
\eqref{eq:reduced_transfer_coefficients}.

The form stress is negative throughout, so the pressure force transfers momentum from the wave to the air. Its magnitude $|\overline\beta|$ decreases as $U_0/c$ increases, so the curves in figure~\ref{fig:exponential_form_stress_comparison}(a) rise towards zero, and at fixed $U_0/c$ it grows with $\delta/\lambda$.
As $c$ and $ak$ are fixed, these trends also hold for the dimensional force.
Both trends follow from the outer integral in
\eqref{eq:reduced_transfer_coefficients}. Increasing $U_0/c$ reduces the
velocity defect $c-\overline U$, and hence $\mathcal I[\overline u]$. A
thicker profile has weaker mean wind over the heights occupied by the
wave-induced response, a larger defect and a larger
$\mathcal I[\overline u]$, so the form stress grows even though the mean
surface shear $U_0/\delta$ falls.

The WKBJ values closely reproduce both the magnitude and the parameter
dependence of the numerical form stress; across the cases shown, the
largest difference from the full finite-difference solution is about
$1.2\%$. For the $R=10^5$ profiles of \S\ref{sec:exp}, the numerical
surface pressure departs from antiphase with the elevation by only
$0.20$--$0.25^\circ$. This small phase shift carries the whole form
stress, because $F_p=ak\Real\{\widehat p_s\}$ in \eqref{eq:form_stress}. For those profiles the WKBJ construction
reproduces the complex pressure to within $1\%$ and its real component to
within $1.6\%$.

Figure~\ref{fig:exponential_form_stress_comparison}(b) extends the
comparison to thin shear layers, $0.05\leq\delta/\lambda\leq0.10$, at
$U_0/c=0.5$ and $R=10^4$. The WKBJ form stress is
within $1.3\%$ of the full numerical value at $\delta/\lambda=0.10$ and
within about $8.5\%$ at $\delta/\lambda=0.05$. 
Since the form stress depends on
the small real part of the surface pressure, it is more sensitive than
the velocity to both the outer WKBJ error and the omitted forcing.

Overall, the WKBJ construction gives excellent agreement with
the numerical form stress for the more slowly varying exponential
profiles, capturing both its magnitude and its dependence on
wind speed and shear-layer thickness.
\subsection{Pressure for the simulation-derived profiles}
\label{sec:cao_pressure}
\label{sec:wrf_pressure}
For all three simulation-derived profiles, the LW--WKBJ pressure is
reconstructed by integrating vertical momentum,
\eqref{eq:WKBJ_reconstruct_p}, downward from $\widehat p(H)=0$.

\begin{figure}[htbp]
\centering
\includegraphics[width=\textwidth]{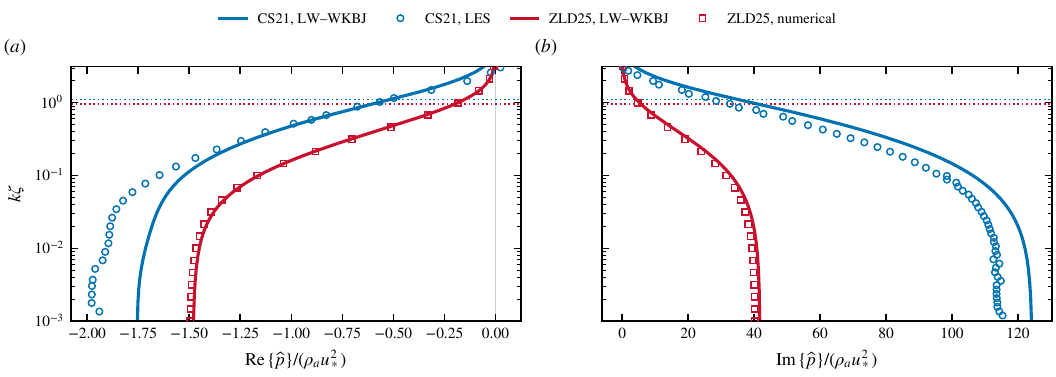}
\caption{Wave-coherent pressure for the simulation-derived means, up to
$k\zeta=\pi$: (a) real and (b) imaginary components. Blue:
\citetalias{caoshen2021} at $c/u_\tau=54.3$, with the two-term LW--WKBJ
(solid) and digitized LES (open circles). Red:
\citetalias{zhang2025windwave} with the two-term LW--WKBJ (solid) and the
numerical solution (open squares).
Dotted lines mark the matching heights, $z_m^+=116.7$ for
\citetalias{caoshen2021} and $z_m^+\approx27.5$ for
\citetalias{zhang2025windwave}. Pressure is reconstructed from integrated
vertical momentum with $\widehat p(H)=0$ and normalized by
$\rho_a u_\ast^2$, with $u_\ast=u_\tau$ for \citetalias{caoshen2021}.}
\label{fig:simulation_pressure}
\end{figure}

Figure~\ref{fig:simulation_pressure} compares the pressure for the mean velocity profiles from both \citetalias{caoshen2021} and \citetalias{zhang2025windwave}.
In both cases the LW--WKBJ captures the dominant imaginary
component and the much smaller real component that determines the form
stress. For \citetalias{caoshen2021}, the errors for the two-term LW--WKBJ are
approximately $10\%$ in each pressure component. A third term reduces the dominant imaginary-pressure discrepancy to
approximately 2\%. For \citet{zhang2025windwave}, adding a third long wave term reduces the complex-pressure discrepancy from $6.35\%$ to $1.00\%$.
\begin{figure}[htbp]
\centering
\includegraphics[width=\textwidth]{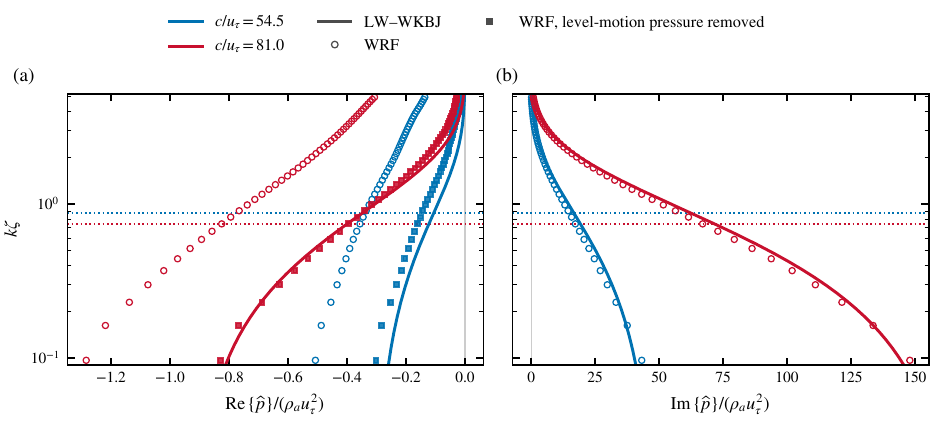}
\caption{Wave-coherent pressure for the wall-modelled WRF LES at
$c/u_\tau=54.5$ (blue) and $c/u_\tau=81.0$ (red): (a) real and
(b) imaginary components. Solid curves show the two-term LW--WKBJ,
reconstructed from integrated vertical momentum with $\widehat p(H)=0$,
and open circles are the WRF harmonics and filled squares the WRF real
pressure with the level-motion pressure of
Appendix~\ref{app:wrf_level_pressure} removed. Dotted lines mark the matching heights.}
\label{fig:wrf_pressure}
\end{figure}
Figure~\ref{fig:wrf_pressure} compares the LW--WKBJ with the WRF harmonics on the resolved levels. The complex pressure discrepancies are below $10\%$ for both wave ages, and retaining a third long-wave term reduces them to below $5\%$. The much smaller real wave-induced pressure component decays more slowly with
height in WRF than in the LW--WKBJ. Appendix~\ref{app:wrf_level_pressure}
identifies an additional pressure contribution associated with
the off-centered acoustic step of the compressible solver.
Off-centering is used to damp vertically propagating sound waves
and maintain numerical stability \citep{klemp2007,zhu2023moving}. When the
computational levels move with the surface wave, however, it also
introduces a small lag in the pressure adjustment. This lag produces
a pressure proportional to the vertical velocity of the levels,
with a magnitude set by the off-centering parameter and the
acoustic time step.

Although this additional pressure is phase-locked to the wave,
it arises from the time discretization and contaminates the
physical wave-induced pressure and form stress. We therefore
subtract the leading numerical contribution derived in
Appendix~\ref{app:wrf_level_pressure}, using the scheme parameters
to determine its magnitude independently of the LW--WKBJ prediction.
After this correction, the WRF real pressure decays with height
as predicted, and its agreement with the LW--WKBJ improves
substantially. The interaction between acoustic damping and moving
computational levels warrants further investigation as compressible
models, including the Energy Research and Forecasting model \citep{lattanzi2025erf}, incorporate surface waves, particularly
because a small numerical pressure can affect the inferred
wave-induced momentum transfer.

\subsection{Form stress for the simulation-derived profiles}
\label{sec:simulation_beta}
For \citetalias{caoshen2021} and \citetalias{zhang2025windwave}, the
LW--WKBJ form stress can also be compared with the numerical solution
of the viscous Orr--Sommerfeld equation. The values for all the
simulation-derived profiles are collected in table~\ref{tab:simulation_beta}.

For \citetalias{caoshen2021}, the nearest digitized LES pressure point
lies at $\zeta/\lambda=2.2\times10^{-4}$. Its real pressure gives the
near-wall estimate $\beta=-25.9$ without extrapolation to the surface, and
the first five digitized points give values between $-26.4$ and $-25.9$.
The LW--WKBJ gives $\beta=-23.4$, and the numerical solution of the
Orr--Sommerfeld equation gives $-22.4$. The two
predictions agree to within $5\%$, so the asymptotic approximation is
accurate. Both are smaller in magnitude than the LES estimate, by about
$10\%$ and $14\%$. Since the numerical solution, which involves no
asymptotic approximation shows a similar difference, it comes from the
reduced equations and the fitted mean, not from the asymptotic
approximation. The \citetalias{zhang2025windwave} archive contains no wave-coherent
pressure, so the LW--WKBJ value $\beta=-11.9$ can be compared only with
the numerical solution of the Orr--Sommerfeld equation, $-12.0$.

The WRF form stress is the resolved surface pressure--slope correlation.
It averages instantaneous pressure--slope products,  extrapolated linearly to the surface from the lowest two mass
levels. With $u_\tau$ from table~\ref{tab:wrf_les}, it gives $\beta=-10.9$
and $-27.5$ for $c/u_\tau=54.5$ and $81.0$. Of these, $-4.1$ and $-9.1$ come
from the level-motion pressure (Appendix~\ref{app:wrf_level_pressure}),
leaving $-6.8$ and $-18.4$. The LW--WKBJ solutions give $-6.1$ and $-18.4$,
within $11\%$ and $1\%$ of these values.

\begin{table}
\centering
\begin{tabular}{lrrrr}
\hline
Case & LW--WKBJ & Numerical & Simulation & Corrected \\
\hline
\citetalias{caoshen2021} & $-23.38$ & $-22.37$ & $-25.87$ & -- \\
\citetalias{zhang2025windwave} & $-11.86$ & $-11.98$ & -- & -- \\
WRF, $c/u_\tau=54.5$ & $-6.09$ & -- & $-10.94$ & $-6.82$ \\
WRF, $c/u_\tau=81.0$ & $-18.36$ & -- & $-27.52$ & $-18.40$ \\
\hline
\end{tabular}
\caption{Form-stress coefficient $\beta$. LW--WKBJ denotes the two-term
LW--WKBJ solution and Numerical is the Finite Difference solution of the viscous
Orr--Sommerfeld problem. The \citetalias{caoshen2021} simulation value
uses the nearest digitized LES pressure point, at
$\zeta/\lambda=2.2\times10^{-4}$; the WRF values are the resolved surface
pressure--slope correlation, and the corrected values remove the
level-motion pressure of Appendix~\ref{app:wrf_level_pressure}. Dashes denote
unavailable comparisons.}
\label{tab:simulation_beta}
\end{table}

Across the simulation-derived profiles, the LW--WKBJ solution therefore
gives the sign and magnitude of the form stress and accurately reproduces results obtained from simulation-derived mean profiles. Geometric forcing has a small effect on the dominant vertical
velocity and complex pressure. The maximum impact of the forcing term was for \citetalias{caoshen2021}, where the difference between the forced and homogeneous solutions for the complex streamwise wave-coherent velocity was about $8\%$.
Together, these results show that the analytical framework accurately predicts form stress for both analytical mean profiles and those
obtained from wall-resolved and wall-modelled simulations.

\section{Conclusions}
\label{sec:conclusions}
We have developed an analytical framework for the wave-coherent
airflow over prescribed fast surface waves. The formulation relates
the mean velocity profile and wave kinematics to the wave-coherent
velocity, pressure, and pressure form stress. The WKBJ construction
for the viscous Orr--Sommerfeld equations separates the response into
an outer Rayleigh mode and a thin viscous mode that enforces the
surface conditions. The outer mode describes how the mean wind
shapes the response over the wave boundary layer, while the viscous
mode supplies the near-surface phase adjustment responsible for
the leading-order form stress. Together, these modes provide a
physical interpretation of the wave-coherent response for both
analytical profiles and simulation-derived turbulent mean flows.

The exponential profile provides a controlled test of the
construction. Where the asymptotic conditions hold, the WKBJ
predictions closely reproduce the numerical velocity and form-stress
trends. For a smooth mean profile held fixed as $R$ increases,
viscous matching supplies the phase correction that produces the
leading $O(R^{-1/2})$ rescaled form stress, while the outer Rayleigh
response determines its coefficient. Applicability requires the absence of a critical level, a positive
Rayleigh potential, separation of the outer and viscous scales,
and sufficiently slow variation of the coefficients governing
each mode over its local vertical scale. The shifted logarithmic profile illustrates the need for a separate
near-wall treatment. Its rapid variation near the surface can
invalidate the WKBJ approximation even without a critical level.
We bridge this region with a long-wave expansion of the inner
Rayleigh solution, matched to the outer WKBJ mode and combined
with the viscous branch. We retain two terms for simplicity,
as these capture the dominant wave-induced response.
 
The LW--WKBJ construction was further assessed using mean profiles
from wall-resolved and wall-modelled simulations. For the
\citetalias{caoshen2021} and WRF cases, the predicted dominant
wave-coherent velocity and pressure responses show excellent
agreement with the LES. The form-stress predictions also compare
favorably with the available near-surface pressure estimates:
the \citetalias{caoshen2021} LW--WKBJ and numerical coefficients
agree closely, although both slightly underestimate the LES estimate,
while the WRF coefficients are comparable to the extrapolated
pressure proxies. For the \citetalias{zhang2025windwave} mean, since wave-coherent DNS fields were not available, the LW--WKBJ was compared to the numerical solution of the viscous Orr--Sommerfeld equation and showed good agreement. These results support the
use of the framework developed here to estimate wave-induced airflow from
observed mean wind profiles, with direct surface-pressure
comparisons providing a further test of its form-stress predictions.

Comparisons of form stress across simulation types need care. The form stress is carried by the phase-shifted components of the wave-coherent response. These are only a few per cent of the dominant motion and are set by the near-surface layer, which is resolved in DNS and wall-resolved LES but represented by a wall model and subgrid stresses in wall-modelled LES.
Differences in this layer that leave the dominant velocity and pressure almost unchanged can therefore change the form stress substantially. At nearly equal wave age, $c/u_{\tau} \approx 54$, the wall-resolved LES of \citetalias{caoshen2021} and the wall-modelled WRF LES performed here (\S\ref{sec:wrf_wall_modelled}) have wave-coherent velocity fields of the same structure once normalized by $akc$, yet their form-stress coefficients are $-25.9$ and $-6.8$, a factor of four apart. 
The LW–WKBJ solution reproduces both, $-23.4$ and $-6.1$, because it takes the near-wall mean and the surface-layer viscosity, molecular or effective, as inputs. 
The leading-order result \eqref{eq:reduced_transfer_coefficients}
identifies the two contributions to this difference. With the decaying
Rayleigh solution computed numerically for each mean, the outer integral
is $\mathcal I[\overline{u}]=0.51$ for \citetalias{caoshen2021} and $0.31$
for WRF, so the stronger mean wind in WRF reduces
$\mathcal I(1+\mathcal I)$ by a factor of $1.9$. The viscous layer
contributes a further factor of $2.3$ through $R^{-1/2}$, with $R=5730$
for \citetalias{caoshen2021} and $R\approx2.9\times10^4$ for WRF, based on
the viscosity at the surface: molecular for \citetalias{caoshen2021} and
effective for WRF. Equation \eqref{eq:reduced_transfer_coefficients} then
gives $\beta=-21.0$ and $-4.9$, a ratio of $4.3$, close to the ratio of
$3.8$ in the simulations. At a given wave age, the form stress therefore
depends on the mean wind over the height of the outer response and on
the viscosity of the surface layer.

The wall-modelled LES comparisons show that the dominant wave-induced velocity structure can be recovered even when the near-surface mean flow is unresolved. This suggests a practical route for applying the 
analytic framework to field observations, where measurements typically begin above the thin viscous layer. Combining measured mean wind profiles and wave characteristics with a suitable near-surface completion would allow the coherent airflow to be estimated at heights where direct phase-resolved measurements are unavailable. The WKBJ diagnostics provide a guide to the conditions under which these estimates are applicable. Future work will test this approach against field measurements such as those reported by \citet{grare2013} and assess its sensitivity to the unresolved near-surface profile. These tests will establish the reliability of using observed mean winds to infer wave-induced velocities, and, subsequently, pressure form stress.

This paper has shown that the airflow over fast waves can be described
analytically, with a priori criteria for its validity, connecting the
mean wind profile directly to the wave-coherent velocity, pressure and
form stress found in wall-resolved, wall-modelled and direct numerical
simulations. The present framework provides a physical basis for developing
wall models that capture both the wave-coherent airflow and its
contribution to momentum transfer. Existing wave-aware wall models
represent wave-dependent drag \citep{aiyer2023,ayala2024}, but
reproducing the associated wave-coherent motion remains a challenge
\citep{kale2026wave}. In the fast-wave regime considered here,
the leading-order form stress
\eqref{eq:reduced_transfer_coefficients} follows from the mean wind
profile, the wave parameters and the surface-layer viscosity.
Together with the analytical velocity and pressure fields, this
relation provides a consistent description of the unresolved
wave-coherent response and the stress it produces. Incorporating
this response into wall-modelled LES offers a pathway to capture
swell-driven momentum transfer without resolving the thin
near-surface viscous layer.

\begin{appen}

\section{Long-wave truncation at fixed matching height}
\label{app:longwave_truncation}
The matched comparisons in \S\ref{sec:test} and
\S\ref{sec:wrf_wall_modelled} retain two terms of the long-wave series.
To isolate this choice, we repeat the construction with $N$ terms,
$\widehat w_{o,N}=\sum_{n=0}^{N-1}k^{2n}\mathcal{W}_n$, for the shifted-logarithmic
example, the \citetalias{caoshen2021} and \citetalias{zhang2025windwave}
means and both WRF cases. The matching height
(table~\ref{tab:longwave_truncation}), mean profile, outer WKBJ mode,
viscous mode and boundary conditions are held fixed; only the mode
amplitudes are recomputed at each $N$ to satisfy the same matching and
wall conditions. Differences are measured against the ten-term solution,
\begin{equation}
\varepsilon_N(\widehat w)=
\frac{\left[\int_0^H|\widehat w_N-\widehat w_{10}|^2\,\mathrm d\zeta\right]^{1/2}}
{\left[\int_0^H|\widehat w_{10}|^2\,\mathrm d\zeta\right]^{1/2}},
\qquad
\varepsilon_N(\beta)=\frac{|\beta_N-\beta_{10}|}{|\beta_{10}|}.
\label{eq:longwave_truncation_error}
\end{equation}
 
\begin{figure}
\centering
\includegraphics[width=\textwidth]{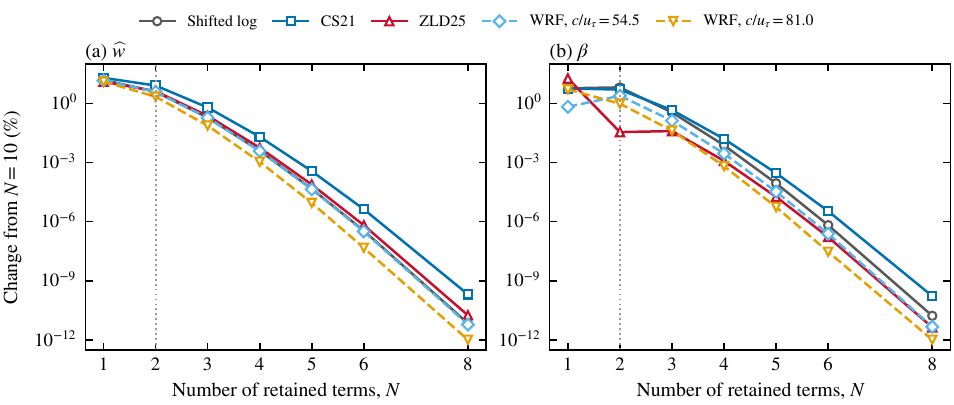}
\caption{Sensitivity to the number of retained long-wave terms at fixed
matching height: for (a) the vertical velocity and
(b) the form-stress coefficient, relative to the ten-term LW--WKBJ. The
dotted line marks the two-term choice of the main text; values below
$10^{-12}\,\%$ are placed at that floor. Matching heights and form-stress
coefficients are listed in table~\ref{tab:longwave_truncation}.}
\label{fig:longwave_truncation}
\end{figure}
 
\begin{table}
\centering
\begin{tabular}{lrrrrr}
\hline
Case & $k\zeta_m$ & $\beta_2$ & $\beta_3$ & $\beta_4$ & $\beta_{10}$ \\
\hline
Shifted log & $0.8860$ & $-0.1160$ & $-0.1093$ & $-0.1089$ & $-0.1089$ \\
\citetalias{caoshen2021} & $1.1061$ & $-23.3837$ & $-22.3719$ & $-22.2727$ & $-22.2692$ \\
\citetalias{zhang2025windwave} & $0.9599$ & $-11.8570$ & $-11.8576$ & $-11.8530$ & $-11.8529$ \\
WRF 54.5 & $0.8732$ & $-6.0907$ & $-5.9540$ & $-5.9462$ & $-5.9460$ \\
WRF 81.0 & $0.7448$ & $-18.3648$ & $-18.1959$ & $-18.1885$ & $-18.1883$ \\
\hline
\end{tabular}
\caption{Fixed matching heights and pressure form-stress coefficients as the total number of long-wave terms increases. The mean, outer WKBJ mode, viscous-mode prescription, and pressure boundary condition are unchanged within each row.}
\label{tab:longwave_truncation}
\end{table}

Two terms differ from the ten-term solution by $2$--$8\%$ in
$\widehat w$ and by up to $6.5\%$ in $\beta$
(figure~\ref{fig:longwave_truncation}). The two-term velocity discrepancy is largest for
\citetalias{caoshen2021}, where the matching height is
$k\zeta_m=1.11$. This is consistent with the growth of the first
omitted term, $k^4\mathcal{W}_2$, with matching height. Retaining
three terms reduces both error measures below $0.7\%$, and four
terms reduce them below $0.03\%$. The horizontal velocity and
pressure show similar convergence.

The converged form-stress coefficients are $-22.27$ for
\citetalias{caoshen2021} and $-11.85$ for
\citetalias{zhang2025windwave}. These agree with the numerical
solutions to within $0.5\%$ and $1.1\%$, respectively
(table~\ref{tab:simulation_beta}). Thus, two terms capture the response to within
a few per cent, while three or four terms substantially reduce
the truncation error without changing the matching height.
\section{Mean-profile fits for the simulation-derived cases}
\label{app:mean_fits}

The mean profiles of \citetalias{caoshen2021} and
\citetalias{zhang2025windwave} are represented by constrained
polynomials in $x_f=\ln(1+z^+)$, where
$z^+=\zeta u_\ast/\nu$:
\begin{equation}
\frac{\overline U(\zeta)}{u_\ast}
=\sum_{j=0}^{6}a_jx_f^j.
\label{eq:app_mean_fit}
\end{equation}
For \citetalias{caoshen2021}, $u_\tau$ replaces $u_\ast$ in
these definitions. Setting $a_0 = 0$ and $a_1 = 1$ enforces $\overline U(0)=0$ and
$\mathrm d\overline U^+/\mathrm dz^+(0)=1$.

The \citetalias{caoshen2021} fit uses 45 mean-velocity samples
with $5.14\leq z^+\leq331$ from the lower half of the channel,
$\zeta/\lambda\leq0.5$, and sets $a_6=0$.
For \citetalias{zhang2025windwave}, heights are measured from
the mean water level, and samples within two wave amplitudes
of the crest are excluded. The remaining 162 samples span
$23.75\leq z^+\leq225$. Since these samples do not resolve the
viscous layer, the fit additionally imposes
$\overline U''(0)=0$, which gives $a_2=1/2$.
The remaining coefficients are determined by unweighted
least squares using only the mean-velocity samples.
Table~\ref{tab:mean_fits} lists the coefficients.
\begin{table}
\centering
\small
\setlength{\tabcolsep}{4pt}
\begin{tabular}{lrrrrr}
\hline
(a) Case & $a_2$ & $a_3$ & $a_4$ & $a_5$ & $a_6$\\
\hline
\citetalias{caoshen2021}
& $0.208418$ & $0.876819$ & $-0.252909$
& $0.0195505$ & $0$\\
\citetalias{zhang2025windwave}
& $1/2$ & $-0.0509172$ & $0.333927$
& $-0.104874$ & $0.00854719$\\
\hline
(b) $c/u_\tau$ & $\kappa$ & $B_{\mathrm{Sp}}$
& $S_0$ ($\mathrm{s^{-1}}$)
& \multicolumn{2}{c}{
$\nu_{\rm eff}(0)$ ($\mathrm{m^2\,s^{-1}}$)}\\
\hline
$54.5$ & $0.322875$ & $9.57245$ & $10.965743$
& \multicolumn{2}{c}{$2.39816\times10^{-3}$}\\
$81.0$ & $0.280135$ & $7.84325$ & $7.002993$
& \multicolumn{2}{c}{$1.69721\times10^{-3}$}\\
\hline
\end{tabular}
\caption{Mean-profile parameters.
(a) Polynomial coefficients in \eqref{eq:app_mean_fit}
for \citetalias{caoshen2021} ($c/u_\tau=54.3$) and
\citetalias{zhang2025windwave} ($c/u_\ast=27.7$).
(b) WRF Spalding completions: $\kappa$ and $B_{\mathrm{Sp}}$
are fitted.}
\label{tab:mean_fits}
\label{tab:wrf_mean_fits}
\end{table}

The WRF means are represented by Equation 
\eqref{eq:wrf_spalding}. The wall value
is prescribed as $\overline U(0)=0$, and the wall slope is
fixed by $S_0=\overline U_{\rm LES}(\zeta_1)/\zeta_1$,
using the first  WRF level.
Only $\kappa$ and $B_{\mathrm{Sp}}$ are fitted, by nonlinear
least squares with equal weight for each retained
mean-velocity sample.

The fits use $u_\tau=0.1621655$ and
$0.1090210\,\mathrm{m\,s^{-1}}$ for the two cases.
Their parameters are listed in
table~\ref{tab:wrf_mean_fits}; $\kappa$ is a fitted shape
parameter rather than a prescribed universal constant.
The relative discrete $L_2$ differences between the fitted
and sampled mean velocities are $1.61\,\%$ and $2.37\,\%$,
respectively. No wave-coherent velocity or pressure data
enter these fits.

\begin{figure}[htbp]
\centering
\includegraphics[width=\textwidth]{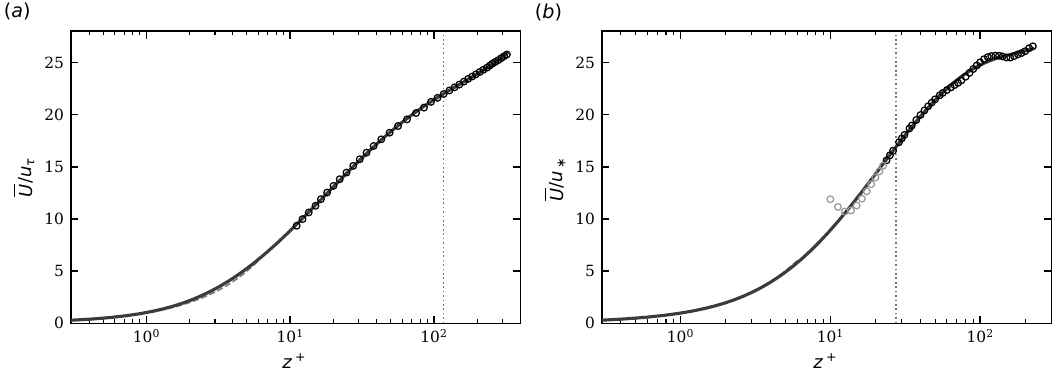}
\caption{Mean streamwise velocity for (a) the lower half of the
wall-resolved LES of \citetalias{caoshen2021}, $c/u_\tau=54.3$, and (b) the
coupled swell DNS of \citetalias{zhang2025windwave}, $c/u_\ast=27.7$, with
heights in (b) measured from the mean water level. Circles denote
simulation data and solid curves the constrained fits
\eqref{eq:app_mean_fit}, which recreate the viscous sublayer,
$\overline U^+=z^+$ (dashed lines), near the wall. Only samples above
$z^+=10$ are shown. In (b), grey circles mark the plane averages within
two wave amplitudes of the crest, which are not used in the fit, and a
logarithmically spaced subset of the retained samples is plotted. Dotted
lines mark the matching heights.}
\label{fig:simulation_mean_fits}
\end{figure}
All mean derivatives used in the WKBJ diagnostics and matched construction are evaluated analytically from these
 fitted profiles. The fits therefore provide consistent
mean velocities, wall conditions and derivatives throughout
the calculation.
\section{Level-motion pressure in the WRF simulations}
\label{app:wrf_level_pressure}

The real part of the WRF pressure decays much more slowly with
height than the LW--WKBJ prediction
(figure~\ref{fig:wrf_pressure}a). Here we show that this excess
is consistent with a numerical pressure contribution arising
from acoustic off-centering as the computational levels move
with the waves.

WRF advances the vertically coupled acoustic variables implicitly
on each acoustic substep $\Delta\tau$. The new and old values
receive weights $(1+\beta_s)/2$ and $(1-\beta_s)/2$, respectively,
where $\beta_s$ is the off-centering parameter
\citep{klemp2007,skamarock2019}. To describe its leading effect,
consider a column state $y$ satisfying $\dot y=\mathsf Ay+f$,
where $\mathsf A$ represents the fast vertical coupling and $f$
the slow forcing, including the prescribed level motion.
To first order in $\Delta\tau$, the off-centered step gives
\begin{equation}
\dot y
=
\mathsf A\left(y+\tfrac12\beta_s\Delta\tau\,\dot y\right)+f
+O(\Delta\tau^2).
\label{eq:wrf_offcentred}
\end{equation}
In the nearly balanced limit, the state in parentheses remains
close to the balanced state $y_s(t)$ of the continuous equations.
Thus $y+\delta t\,\dot y\approx y_s$, with
$\delta t=\beta_s\Delta\tau/2$, or equivalently
$y(t)\approx y_s(t-\delta t)$. The computed balance therefore
lags the prescribed motion of the levels.

For the hydrostatic part of the balance, a level displaced
vertically by $h$ experiences a pressure change
$\partial_t p_s=-\rho g\,\partial_t h$.
The lag leaves a residual
$p_{\mathrm{num}}\approx-\delta t\,\partial_t p_s$, giving
\begin{equation}
p_{\mathrm{num}}
=
\rho g\,\delta t\,\partial_t h,
\qquad
\frac{\widehat p_{\mathrm{num}}}{\rho}
=
-\mathrm i g\,\delta t\,kc\,\widehat h.
\label{eq:wrf_level_pressure}
\end{equation}
For the WRF mapping,
$\widehat h=(1-\zeta/H)\widehat\eta$.
Under the Fourier convention used here, $\widehat\eta$ is
imaginary, so $\widehat p_{\mathrm{num}}$ is real.
At this order, the residual therefore changes the pressure
component responsible for form stress, while leaving
$\Imag\{\widehat p\}$ unchanged. Its surface contribution to
the form-stress coefficient is
$\beta_{\mathrm{num}}=-g\,\delta t\,c/u_\tau^2$.

The scheme parameters provide a prediction that can be checked
against the pressure profiles. The simulations use $\beta_s=1$
and $\Delta\tau=\Delta t/4=2.5$\,ms, giving
$\delta t=1.25$\,ms. Independently, we estimate $\delta t$ by
fitting \eqref{eq:wrf_level_pressure} to
$\Real\{\widehat p\}$ above $k\zeta=6.5$, where the physical
wave-coherent pressure is expected to be negligible.
The fitted delays are $1.24$ and $1.18$\,ms for
$c/u_\tau=54.5$ and $81.0$, respectively, and vary by less
than $15\%$ between quarters of the averaging window.
The slowly decaying residual is also well described by the
predicted height dependence, $1-\zeta/H$.
The agreement between the predicted and fitted delays supports
the interpretation of this residual as a contribution from
acoustic off-centering.
Using the scheme-based value $\delta t=1.25$\,ms gives
$\beta_{\mathrm{num}}=-4.1$ and $-9.1$, respectively,
accounting for about a third of the WRF pressure--slope
estimates in table~\ref{tab:simulation_beta}.
Figure~\ref{fig:wrf_pressure}(a) shows the real pressure after
subtracting this contribution.

This leading-order model identifies a numerical contribution
to the WRF form stress that scales with $\beta_s\Delta\tau$.
It predicts that the contribution vanishes for a centered
acoustic step ($\beta_s=0$) or stationary computational levels.
The specific mechanism is absent from incompressible
formulations such as that of \citetalias{caoshen2021}, which
do not use acoustic time splitting.
\end{appen}\clearpage

\begin{bmhead}[Acknowledgements]{AA thanks Professor K. Subramanian for insightful discussions and helpful comments on the manuscript.}
\end{bmhead}
\begin{bmhead}[Conflict of interest]
    The author declares no competing interests.
\end{bmhead}

\begin{bmhead}[Funding]
    Research reported in this publication was supported by an Early-Career Research Fellowship from the Gulf Research Program of the National Academies of Sciences, Engineering, and Medicine. The content is solely the responsibility of the author and does not necessarily represent the official views of the Gulf Research Program of the National Academies of Sciences, Engineering, and Medicine.
\end{bmhead}

\begin{bmhead}[Use of AI tools]
 The author used generative AI assistants (Claude and Codex) for refining the numerical finite difference solver and editing the text. The author checked all the numerical
results, figures, and citations.   
\end{bmhead}

\bibliographystyle{jfm}
\bibliography{refs,refs_additions}

@article{zhu2023moving,
  title={A moving-wave implementation in {WRF} to study the impact of surface water waves on the atmospheric boundary layer},
  author={Zhu, Peiyun and Li, Tianyi and Mirocha, Jeffrey D and Arthur, Robert S and Wu, Zhao and Fringer, Oliver B},
  journal={Monthly Weather Review},
  volume={151},
  number={11},
  pages={2883--2903},
  year={2023},
  publisher={American Meteorological Society}
}

@article{phillips1957,
  author  = {Phillips, O. M.},
  title   = {On the generation of waves by turbulent wind},
  journal = {J. Fluid Mech.},
  year    = {1957},
  volume  = {2},
  number  = {5},
  pages   = {417--445},
  doi     = {10.1017/S0022112057000233},
}

@article{miles1957,
  author  = {Miles, John W.},
  title   = {On the generation of surface waves by shear flows},
  journal = {J. Fluid Mech.},
  year    = {1957},
  volume  = {3},
  number  = {2},
  pages   = {185--204},
  doi     = {10.1017/S0022112057000567},
}

@article{miles1959,
  author  = {Miles, John W.},
  title   = {On the generation of surface waves by shear flows. {Part} 2},
  journal = {J. Fluid Mech.},
  year    = {1959},
  volume  = {6},
  number  = {4},
  pages   = {568--582},
  doi     = {10.1017/S0022112059000830},
}

@article{benjamin1959,
  author  = {Benjamin, T. Brooke},
  title   = {Shearing flow over a wavy boundary},
  journal = {J. Fluid Mech.},
  year    = {1959},
  volume  = {6},
  number  = {2},
  pages   = {161--205},
  doi     = {10.1017/S0022112059000568},
}

@article{miles1962,
  author  = {Miles, John W.},
  title   = {On the generation of surface waves by shear flows. {Part} 4},
  journal = {J. Fluid Mech.},
  year    = {1962},
  volume  = {13},
  number  = {3},
  pages   = {433--448},
  doi     = {10.1017/S0022112062000828},
}

@article{vangastel1985,
  author  = {Van Gastel, K. and Janssen, P. A. E. M. and Komen, G. J.},
  title   = {On phase velocity and growth rate of wind-induced
             gravity-capillary waves},
  journal = {J. Fluid Mech.},
  year    = {1985},
  volume  = {161},
  pages   = {199--216},
  doi     = {10.1017/S0022112085002889},
}

@article{miles1993,
  author  = {Miles, John W.},
  title   = {Surface-wave generation revisited},
  journal = {J. Fluid Mech.},
  year    = {1993},
  volume  = {256},
  pages   = {427--441},
  doi     = {10.1017/S0022112093002836},
}

@article{cohenbelcher1999,
  author  = {Cohen, J. E. and Belcher, S. E.},
  title   = {Turbulent shear flow over fast-moving waves},
  journal = {J. Fluid Mech.},
  year    = {1999},
  volume  = {386},
  pages   = {345--371},
  doi     = {10.1017/S0022112099004383},
}

@article{bonfils2022,
  author  = {Bonfils, A. F. and Mitra, Dhrubaditya and Moon, W. and
             Wettlaufer, J. S.},
  title   = {Asymptotic interpretation of the {Miles} mechanism of
             wind-wave instability},
  journal = {J. Fluid Mech.},
  year    = {2022},
  volume  = {944},
  pages   = {A8},
  doi     = {10.1017/jfm.2022.441},
}

@article{bonfils2023,
  author  = {Bonfils, A. F. and Mitra, Dhrubaditya and Moon, W. and
             Wettlaufer, J. S.},
  title   = {Flow-driven interfacial waves: an inviscid asymptotic study},
  journal = {J. Fluid Mech.},
  year    = {2023},
  volume  = {976},
  pages   = {A19},
  doi     = {10.1017/jfm.2023.906},
}

@book{drazinreid2004,
  author    = {Drazin, P. G. and Reid, W. H.},
  title     = {Hydrodynamic Stability},
  edition   = {2nd},
  publisher = {Cambridge University Press},
  year      = {2004},
  doi       = {10.1017/CBO9780511616938},
}

@article{hristov2003,
  author  = {Hristov, T. S. and Miller, S. D. and Friehe, C. A.},
  title   = {Dynamical coupling of wind and ocean waves through
             wave-induced air flow},
  journal = {Nature},
  year    = {2003},
  volume  = {422},
  pages   = {55--58},
  doi     = {10.1038/nature01382},
}

@article{grare2013,
  author  = {Grare, L. and Lenain, L. and Melville, W. K.},
  title   = {Wave-coherent airflow and critical layers over ocean waves},
  journal = {J. Phys. Oceanogr.},
  year    = {2013},
  volume  = {43},
  number  = {10},
  pages   = {2156--2172},
  doi     = {10.1175/JPO-D-13-056.1},
}

@article{carpenter2022,
  author  = {Carpenter, J. R. and Buckley, M. P. and Veron, F.},
  title   = {Evidence of the critical layer mechanism in growing wind waves},
  journal = {J. Fluid Mech.},
  year    = {2022},
  volume  = {948},
  pages   = {A26},
  doi     = {10.1017/jfm.2022.714},
}

@article{klemp2007,
  author  = {Klemp, J. B. and Skamarock, W. C. and Dudhia, J.},
  title   = {Conservative split-explicit time integration methods for the
             compressible nonhydrostatic equations},
  journal = {Mon. Weather Rev.},
  year    = {2007},
  volume  = {135},
  number  = {8},
  pages   = {2897--2913},
  doi     = {10.1175/MWR3440.1},
}

@article{smedman1994,
  author  = {Smedman, A.-S. and Tjernstr{\"o}m, M. and H{\"o}gstr{\"o}m, U.},
  title   = {The near-neutral marine atmospheric boundary layer with no
             surface shearing stress: a case study},
  journal = {J. Atmos. Sci.},
  year    = {1994},
  volume  = {51},
  number  = {23},
  pages   = {3399--3411},
  doi     = {10.1175/1520-0469(1994)051<3399:TNNMAB>2.0.CO;2},
}

@article{grachevfairall2001,
  author  = {Grachev, A. A. and Fairall, C. W.},
  title   = {Upward momentum transfer in the marine boundary layer},
  journal = {J. Phys. Oceanogr.},
  year    = {2001},
  volume  = {31},
  number  = {7},
  pages   = {1698--1711},
  doi     = {10.1175/1520-0485(2001)031<1698:UMTITM>2.0.CO;2},
}

@article{sullivan2008,
  author  = {Sullivan, P. P. and Edson, J. B. and Hristov, T. and
             McWilliams, J. C.},
  title   = {Large-eddy simulations and observations of atmospheric marine
             boundary layers above nonequilibrium surface waves},
  journal = {J. Atmos. Sci.},
  year    = {2008},
  volume  = {65},
  number  = {4},
  pages   = {1225--1245},
  doi     = {10.1175/2007JAS2427.1},
}

@article{hanleybelcher2010,
  title={A global climatology of wind--wave interaction},
  author={Hanley, Kirsty E and Belcher, Stephen E and Sullivan, Peter P},
  journal={Journal of physical oceanography},
  volume={40},
  number={6},
  pages={1263--1282},
  year={2010}
}

@article{akervikvartdal2019,
  author  = {{\AA}kervik, E. and Vartdal, M.},
  title   = {The role of wave kinematics in turbulent flow over waves},
  journal = {J. Fluid Mech.},
  year    = {2019},
  volume  = {880},
  pages   = {890--915},
  doi     = {10.1017/jfm.2019.708},
}

@article{caodengshen2020,
  author  = {Cao, T. and Deng, B.-Q. and Shen, L.},
  title   = {A simulation-based mechanistic study of turbulent wind blowing
             over opposing water waves},
  journal = {J. Fluid Mech.},
  year    = {2020},
  volume  = {901},
  pages   = {A27},
  doi     = {10.1017/jfm.2020.591},
}

@article{caoshen2021,
  author  = {Cao, T. and Shen, L.},
  title   = {A numerical and theoretical study of wind over
             fast-propagating water waves},
  journal = {J. Fluid Mech.},
  year    = {2021},
  volume  = {919},
  pages   = {A38},
  doi     = {10.1017/jfm.2021.416},
}

@article{narasimhan2026,
  author  = {Narasimhan, Ghanesh and Deskos, Georgios and Ren, Ziyan and
             Shen, Lian},
  title   = {Revisiting the validity of eddy viscosity models for predicting
             airflow over water waves},
  journal = {J. Fluid Mech.},
  year    = {2026},
  volume  = {1026},
  pages   = {A22},
  doi     = {10.1017/jfm.2025.11015},
}

@article{dore1978,
  author  = {Dore, B. D.},
  title   = {Some effects of the air-water interface on gravity waves},
  journal = {Geophys. Astrophys. Fluid Dyn.},
  year    = {1978},
  volume  = {10},
  number  = {1},
  pages   = {215--230},
  doi     = {10.1080/03091927808242638},
}

@article{perignon2014,
  author  = {P{\'e}rignon, Yves and Ardhuin, Fabrice and Cathelain, Marie and Robert, Marc},
  title   = {Swell dissipation by induced atmospheric shear stress},
  journal = {J. Geophys. Res. Oceans},
  year    = {2014},
  volume  = {119},
  number  = {10},
  pages   = {6622--6630},
  doi     = {10.1002/2014JC009896},
}

@article{Semedo2009,
  author  = {Semedo, Alvaro and S{\ae}tra, {\O}yvind and Rutgersson, Anna and Kahma, Kimmo K. and Pettersson, Heidi},
  title   = {Wave-induced wind in the marine boundary layer},
  journal = {J. Atmos. Sci.},
  year    = {2009},
  volume  = {66},
  number  = {8},
  pages   = {2256--2271},
  doi     = {10.1175/2009JAS3018.1},
}

@article{kawai1979,
  author  = {Kawai, Sanshiro},
  title   = {Generation of initial wavelets by instability of a coupled shear flow and their evolution to wind waves},
  journal = {J. Fluid Mech.},
  year    = {1979},
  volume  = {93},
  number  = {4},
  pages   = {661--703},
  doi     = {10.1017/S002211207900197X},
}

@article{valenzuela1976,
  author  = {Valenzuela, Gaspar R.},
  title   = {The growth of gravity-capillary waves in a coupled shear flow},
  journal = {J. Fluid Mech.},
  year    = {1976},
  volume  = {76},
  number  = {2},
  pages   = {229--250},
  doi     = {10.1017/S0022112076000608},
}

@article{wu2016,
  author  = {Wu, Lichuan and Rutgersson, Anna and Sahl{\'e}e, Erik and
             Lars{\'e}n, Xiaoli Guo},
  title   = {Swell impact on wind stress and atmospheric mixing in a regional
             coupled atmosphere--wave model},
  journal = {J. Geophys. Res. Oceans},
  year    = {2016},
  volume  = {121},
  number  = {7},
  pages   = {4633--4648},
  doi     = {10.1002/2015JC011576},
}

@article{kumar2026,
  author  = {Kumar, Krishanu and Shemer, Lev},
  title   = {Shear flow instability analysis of young wind waves: coupled
             air--water {O}rr--{S}ommerfeld framework guided by experiment},
  journal = {J. Fluid Mech.},
  year    = {2026},
  volume  = {1030},
  pages   = {A39},
  doi     = {10.1017/jfm.2026.11250},
}

@article{zhang2025windwave,
  author  = {Zhang, Jinlong and Liu, Weijian and Dong, Yuhong},
  title   = {Impact of Wind-Wave Coupling on Turbulence and Air-Sea Fluxes:
             Insights From Direct Numerical Simulations},
  journal = {Geophysical Research Letters},
  year    = {2025},
  volume  = {52},
  pages   = {e2025GL117290},
  doi     = {10.1029/2025GL117290},
}

@article{ning2024parameterization,
  title={Parameterization of wave-induced stress in large-eddy simulations of the marine atmospheric boundary layer},
  author={Ning, Xu and Paskyabi, Mostafa Bakhoday},
  journal={Journal of Geophysical Research: Oceans},
  volume={129},
  number={9},
  pages={e2023JC020722},
  year={2024},
  publisher={Wiley Online Library}
}

@article{lattanzi2025erf,
  title={{ERF}: Energy research and forecasting model},
  author={Lattanzi, Aaron and Almgren, Ann and Quon, Eliot and Natarajan, Mahesh and Kosovic, Branko and Mirocha, Jeffrey and Perry, Bruce and Wiersema, David and Willcox, Donald and Yuan, Xingqiu and others},
  journal={Journal of Advances in Modeling Earth Systems},
  volume={17},
  number={11},
  pages={e2024MS004884},
  year={2025},
  publisher={Wiley Online Library}
}

@article{kale2026wave,
  title={Wave-phase-aware model implementation in {WRF-LES v3. 8.1} for turbulence-resolving simulations of turbulent flow over monochromatic waves},
  author={Kale, Baris and Asadbeigi, Mohammadreza and Aiyer, Aditya K and Howland, Michael F},
  journal={EGUsphere},
  volume={2026},
  pages={1--26},
  year={2026},
  publisher={Copernicus Publications, G{\"o}ttingen, Germany}
}

@article{drazinhoward1962,
  author  = {Drazin, P. G. and Howard, L. N.},
  title   = {The instability to long waves of unbounded parallel inviscid
             flow},
  journal = {J. Fluid Mech.},
  year    = {1962},
  volume  = {14},
  number  = {2},
  pages   = {257--283},
  doi     = {10.1017/S0022112062001238},
}

@book{benderorszag1999,
  author    = {Bender, C. M. and Orszag, S. A.},
  title     = {Advanced Mathematical Methods for Scientists and Engineers
               {I}: Asymptotic Methods and Perturbation Theory},
  publisher = {Springer},
  address   = {New York},
  year      = {1999},
}

@article{youngwolfe2014,
  author  = {Young, W. R. and Wolfe, C. L.},
  title   = {Generation of surface waves by shear-flow instability},
  journal = {J. Fluid Mech.},
  year    = {2014},
  volume  = {739},
  pages   = {276--307},
  doi     = {10.1017/jfm.2013.617},
}

@article{carpenter2017,
  author  = {Carpenter, J. R. and Guha, A. and Heifetz, E.},
  title   = {A physical interpretation of the wind-wave instability as
             interacting waves},
  journal = {J. Phys. Oceanogr.},
  year    = {2017},
  volume  = {47},
  number  = {6},
  pages   = {1441--1455},
  doi     = {10.1175/JPO-D-16-0206.1},
}

@article{muppirala2026,
  author  = {Muppirala, H. and Patibandla, R. and Roy, A.},
  title   = {On the generation of free-surface waves by instabilities in
             quadratic shear flows},
  journal = {J. Fluid Mech.},
  year    = {2026},
  volume  = {1037},
  pages   = {A68},
  doi     = {10.1017/jfm.2026.11705},
}

@article{townsend1972,
  author  = {Townsend, A. A.},
  title   = {Flow in a deep turbulent boundary layer over a surface
             distorted by water waves},
  journal = {J. Fluid Mech.},
  year    = {1972},
  volume  = {55},
  number  = {4},
  pages   = {719--735},
  doi     = {10.1017/S0022112072002101},
}

@article{plant1982,
  author  = {Plant, W. J.},
  title   = {A relationship between wind stress and wave slope},
  journal = {J. Geophys. Res.},
  year    = {1982},
  volume  = {87},
  number  = {C3},
  pages   = {1961--1967},
  doi     = {10.1029/JC087iC03p01961},
}

@article{janssen1991,
  author  = {Janssen, P. A. E. M.},
  title   = {Quasi-linear theory of wind-wave generation applied to wave
             forecasting},
  journal = {J. Phys. Oceanogr.},
  year    = {1991},
  volume  = {21},
  number  = {11},
  pages   = {1631--1642},
  doi     = {10.1175/1520-0485(1991)021<1631:QLTOWW>2.0.CO;2},
}

@article{sullivan2000,
  author  = {Sullivan, P. P. and McWilliams, J. C. and Moeng, C.-H.},
  title   = {Simulation of turbulent flow over idealized water waves},
  journal = {J. Fluid Mech.},
  year    = {2000},
  volume  = {404},
  pages   = {47--85},
  doi     = {10.1017/S0022112099006965},
}

@article{wu2022,
  author  = {Wu, J. and Popinet, S. and Deike, L.},
  title   = {Revisiting wind wave growth with fully coupled direct
             numerical simulations},
  journal = {J. Fluid Mech.},
  year    = {2022},
  volume  = {951},
  pages   = {A18},
  doi     = {10.1017/jfm.2022.822},
}

@article{kudryavtsevmakin2004,
  author  = {Kudryavtsev, V. N. and Makin, V. K.},
  title   = {Impact of swell on the marine atmospheric boundary layer},
  journal = {J. Phys. Oceanogr.},
  year    = {2004},
  volume  = {34},
  number  = {4},
  pages   = {934--949},
  doi     = {10.1175/1520-0485(2004)034<0934:IOSOTM>2.0.CO;2},
}

@article{hanleybelcher2008,
  author  = {Hanley, K. E. and Belcher, S. E.},
  title   = {Wave-driven wind jets in the marine atmospheric boundary
             layer},
  journal = {J. Atmos. Sci.},
  year    = {2008},
  volume  = {65},
  number  = {8},
  pages   = {2646--2660},
  doi     = {10.1175/2007JAS2562.1},
}

@article{ardhuin2009,
  author  = {Ardhuin, F. and Chapron, B. and Collard, F.},
  title   = {Observation of swell dissipation across oceans},
  journal = {Geophys. Res. Lett.},
  year    = {2009},
  volume  = {36},
  number  = {6},
  pages   = {L06607},
  doi     = {10.1029/2008GL037030},
}

@article{smagorinsky1963,
  author  = {Smagorinsky, J.},
  title   = {General circulation experiments with the primitive
             equations: {I}. {T}he basic experiment},
  journal = {Mon. Weather Rev.},
  year    = {1963},
  volume  = {91},
  number  = {3},
  pages   = {99--164},
  doi     = {10.1175/1520-0493(1963)091<0099:GCEWTP>2.3.CO;2},
}

@article{spalding1961,
  author  = {Spalding, D. B.},
  title   = {A single formula for the ``law of the wall''},
  journal = {J. Appl. Mech.},
  year    = {1961},
  volume  = {28},
  number  = {3},
  pages   = {455--458},
  doi     = {10.1115/1.3641728},
}

@techreport{skamarock2019,
  author      = {Skamarock, W. C. and Klemp, J. B. and Dudhia, J. and
                 Gill, D. O. and Liu, Z. and Berner, J. and Wang, W. and
                 Powers, J. G. and Duda, M. G. and Barker, D. M. and
                 Huang, X.-Y.},
  title       = {A description of the {A}dvanced {R}esearch {WRF} model
                 version 4},
  institution = {National Center for Atmospheric Research},
  type        = {NCAR Tech. Note},
  number      = {NCAR/TN-556+STR},
  year        = {2019},
  doi         = {10.5065/1dfh-6p97},
}

@article{aiyer2023,
  author  = {Aiyer, A. K. and Deike, L. and Mueller, M. E.},
  title   = {A sea surface--based drag model for large-eddy simulation of
             wind--wave interaction},
  journal = {J. Atmos. Sci.},
  year    = {2023},
  volume  = {80},
  number  = {1},
  pages   = {49--62},
  doi     = {10.1175/JAS-D-21-0329.1},
}

@article{aiyer2024,
  author  = {Aiyer, A. K. and Deike, L. and Mueller, M. E.},
  title   = {A dynamic wall modeling approach for large eddy simulation of
             offshore wind farms in realistic oceanic conditions},
  journal = {J. Renew. Sustain. Energy},
  year    = {2024},
  volume  = {16},
  number  = {1},
  pages   = {013305},
  doi     = {10.1063/5.0159019},
}

@article{ayala2024,
  author  = {Ayala, M. and Sadek, Z. and Fer{\v{c}}{\'a}k, O. and Cal, R. B.
             and Gayme, D. F. and Meneveau, C.},
  title   = {A moving surface drag model for {LES} of wind over waves},
  journal = {Boundary-Layer Meteorol.},
  year    = {2024},
  volume  = {190},
  number  = {10},
  pages   = {39},
  doi     = {10.1007/s10546-024-00884-8},
}

@article{wu2024review,
  title={A review of surface swell waves and their role in air--sea interactions},
  author={Wu, Lichuan and Sahl{\'e}e, Erik and Nilsson, Erik and Rutgersson, Anna},
  journal={Ocean Modelling},
  volume={190},
  pages={102397},
  year={2024},
  publisher={Elsevier}
}

@article{lishen2022,
  author  = {Li, T. and Shen, L.},
  title   = {The principal stage in wind-wave generation},
  journal = {J. Fluid Mech.},
  year    = {2022},
  volume  = {934},
  pages   = {A41},
  doi     = {10.1017/jfm.2021.1153},
}

@article{lishen2025,
  author  = {Li, T. and Shen, L.},
  title   = {A theoretical study of the upper bound of surface elevation
             variance in the {P}hillips initial stage during wind-wave
             generation},
  journal = {J. Fluid Mech.},
  year    = {2025},
  volume  = {1006},
  pages   = {A8},
  doi     = {10.1017/jfm.2025.13},
}

@article{yangshen2010,
  author  = {Yang, D. and Shen, L.},
  title   = {Direct-simulation-based study of turbulent flow over various
             waving boundaries},
  journal = {J. Fluid Mech.},
  year    = {2010},
  volume  = {650},
  pages   = {131--180},
  doi     = {10.1017/S0022112009993557},
}

@article{haoshen2019,
  author  = {Hao, X. and Shen, L.},
  title   = {Wind--wave coupling study using {LES} of wind and phase-resolved
             simulation of nonlinear waves},
  journal = {J. Fluid Mech.},
  year    = {2019},
  volume  = {874},
  pages   = {391--425},
  doi     = {10.1017/jfm.2019.444},
}

@article{zhangdongshen2024,
  author  = {Zhang, J. and Dong, Y. and Shen, L.},
  title   = {Characteristics of the dissimilar turbulent transport processes
             of heat and momentum during wind--wave dynamical interactions},
  journal = {J. Geophys. Res. Oceans},
  year    = {2024},
  volume  = {129},
  number  = {12},
  pages   = {e2024JC021320},
  doi     = {10.1029/2024JC021320},
}

@article{kadam2023,
  author  = {Kadam, Y. and Patibandla, R. and Roy, A.},
  title   = {Wind-generated waves on a water layer of finite depth},
  journal = {J. Fluid Mech.},
  year    = {2023},
  volume  = {967},
  pages   = {A12},
  doi     = {10.1017/jfm.2023.483},
}

@article{kumarravichandran2026,
  author  = {Kumar, A. and Ravichandran, S. and Dasgupta, R.},
  title   = {Waves in a shear flow: transition between the {KH}, {H}olmboe
             and {M}iles instability},
  journal = {J. Fluid Mech.},
  year    = {2026},
  volume  = {1036},
  pages   = {A10},
  doi     = {10.1017/jfm.2026.11588},
}

\end{document}